\documentclass{aastex}               
\usepackage{emulateapj5}             
\newcommand{\fuse}{{\it FUSE\/}}
\newcommand{\kms}{${\rm km\,s}^{-1}$}
\newcommand{\cii}{\ion{C}{2}}
\newcommand{\ciii}{\ion{C}{3}}
\newcommand{\civ}{\ion{C}{4}}
\newcommand{\siiii}{\ion{Si}{3}}
\newcommand{\siiv}{\ion{Si}{4}}
\newcommand{\siv}{\ion{S}{4}}
\newcommand{\svi}{\ion{S}{6}}
\newcommand{\piv}{\ion{P}{4}}
\newcommand{\pv}{\ion{P}{5}}

\newcommand{\nii}{\ion{N}{2}}
\newcommand{\niii}{\ion{N}{3}}
\newcommand{\niv}{\ion{N}{4}}
\newcommand{\nv}{\ion{N}{5}}

\newcommand{\oiii}{\ion{O}{3}}
\newcommand{\ovi}{\ion{O}{6}}
\newcommand{\heii}{\ion{He}{2}}

\slugcomment{Submitted to Astrophysical Journal Supplement Series}
\slugcomment{Draft of \today}
\shorttitle{ {\fuse} Atlas of Galactic OB Stars}
\shortauthors{Pellerin et al.}

\begin{document}

\title{ An Atlas of Galactic OB Spectra Observed with the \\
       {\it Far Ultraviolet Spectroscopic Explorer}
       	 \footnote{Based on observations made with the NASA-CNES-CSA Far
       	 Ultraviolet Spectroscopic Explorer. FUSE is operated for NASA by 
	 the Johns Hopkins University under NASA contract NAS5-32985.} }    

\author{ Anne Pellerin\altaffilmark{2}, 
	Alex  W. Fullerton\altaffilmark{3,4},
	Carmelle Robert\altaffilmark{2}, 
        J. Christopher Howk\altaffilmark{4},
	John  B. Hutchings\altaffilmark{5},  
	Nolan R. Walborn\altaffilmark{6},
	Luciana  Bianchi\altaffilmark{4,7},
	Paul  A. Crowther\altaffilmark{8}, and
	George  Sonneborn\altaffilmark{9}  }

\altaffiltext{2}{D\'epartement de physique, de génie physique et d'optique,
		 Universit\'e Laval and Observatoire du mont M\'egantic, 
		 Qu\'ebec, QC, G1K 7P4, 
		 Canada. }
		 \email{apelleri@phy.ulaval.ca}

\altaffiltext{3}{Department of Physics \& Astronomy, 
                 University of Victoria, 
                 P.O. Box 3055, 
		 Victoria, BC, V8W 3P6, Canada.}

\altaffiltext{4}{Department of Physics \& Astronomy, 
                 The Johns Hopkins University, 
		 3400 N. Charles Street, 
		 Baltimore, MD 21218, USA.}

\altaffiltext{5}{Herzberg Institute of Astrophysics,
                 National Research Council of Canada,
                 5071 West Saanich Road,
                 Victoria, BC, V8X 4M6,
                 Canada.}

\altaffiltext{6}{Space Telescope Science Institute,
		 3700 San Martin Drive,
		 Baltimore, MD 21218,USA.}

\altaffiltext{7}{On leave from Osservatorio Astronomico di Torino, Italy.}

\altaffiltext{8}{Department of Physics \& Astronomy,
                 University College London,        
		 Gower Street,
		 London, WC1E 6BT,
		 U.K.}
		 
\altaffiltext{9}{Laboratory for Astronomy and Solar Physics,
		 NASA Goddard Space Flight Center,
		 Code 681, Greenbelt,
    	         MD 20771,USA.}

\begin{abstract}

An atlas of far-ultraviolet spectra of 45 Galactic OB stars observed
with the {\it Far Ultraviolet Spectroscopic Explorer} is presented.
The atlas covers the wavelength region between 912 and 1185~{\AA} with
an effective spectral resolution of 0.12~{\AA}. Systematic trends in the
morphology and strength of stellar features are discussed. Particular
attention is drawn to the variations of the {\ciii~$\lambda$1176},
{\siv~$\lambda\lambda$1063, 1073}, and {\pv~$\lambda\lambda$1118, 1128}
line profiles as a function of temperature and luminosity class; and the
lack of a luminosity dependence associated with
{\ovi~$\lambda\lambda$1032, 1038}.
Numerous interstellar lines are also identified.

\end{abstract}


\section{Introduction}

The far-ultraviolet (FUV; 900--1200~{\AA}) region of the spectrum contains
an enormous number of spectral features attributable to resonance lines and
transitions between excited states. These transitions are due to a variety
of atomic and molecular species, which include elements that are cosmically
abundant and some that are comparatively rare. Collectively, these lines
diagnose a wide range of ionization and excitation conditions, and can
therefore provide extremely detailed information about the physical
conditions that exist in astrophysically interesting environments like
stellar atmospheres, the interstellar medium (ISM), and the intergalactic
medium (IGM).

Unfortunately, observations in the FUV are also very challenging from a
technical standpoint, both because of the low reflectivity of optical
surfaces at such short wavelengths and the need for high spectral
resolution to minimize confusion from line blending. Consequently, despite
the many scientific incentives, the FUV window has been underutilized.
Before 1999, the main forays into the FUV were limited to the 
{\it Copernicus} satellite 
{\citep[1972--1981;][]{rog73}} 
and a series of comparatively short duration shuttle-based missions:
the Hopkins Ultraviolet Telescope 
{\citep[HUT;][]{dav92}} 
which flew on the {\it Astro-1} and {\it Astro-2} missions; and
the Interstellar Medium Absorption Profile Spectrometer 
(IMAPS; Jenkins, Reale, \& Zucchino 1996),
the Berkeley Extreme and Far Ultraviolet Spectrometer 
{\citep[BEFS;][]{hur98}},
and the T{\"u}bingen Echelle Spectrograph 
{\citep[TUES;][]{barn99}}, 
which flew on the {\it ORFEUS-SPAS I} and {\it II} missions.

The launch of the {\it Far Ultraviolet Spectroscopic Explorer} ({\fuse}) in 
1999 June rectified this situation. Since then, {\fuse} has provided
routine access to the entire FUV waveband with high spectral resolution
and exceptional sensitivity, which in turn has permitted observations of a
substantially larger target pool than was available to {\it Copernicus}
(due to sensitivity limitations) or the shuttle-based missions (due to
time constraints). In particular, spectra of more than 200 Galactic OB-type
stars covering most spectral types\footnote[10]{See \citet{wal02} for a
description of the criteria that define the new O2 spectral class.} and
luminosity classes between O2 and B9 have been obtained as part of the
various programs implemented by the Principal Investigator (PI) Team.

We have selected a subset of these objects for presentation in a 
high-resolution FUV spectral atlas. The principal aim of this atlas is to
show the general behavior of the most prominent stellar lines as a function
of temperature and luminosity class, and also to illustrate the rich
diversity of the interstellar spectrum. This work has several immediate
applications, such as line identification, the characterization of
hot-star winds, FUV extinction, population synthesis, and the
interpretation of young stellar populations in distant galaxies. More
fundamentally, we hope that this atlas will help familiarize researchers
with the FUV region of the spectrum. It complements existing atlases based
on spectra obtained with {\it Copernicus\/} {\citep[e.g.][]{snow76,
snow77, wal96}} both by including the spectral region between 910 and
1000\AA\ and by enlarging the sample of objects earlier than B3, and serves
as a companion to the {\fuse} atlas of OB-type stars in the Magellanic
Clouds \citet{mcatlas}.

The remainder of this paper is organized as follows. The observational
material and data processing are summarized in {\S 2}, while {\S 3}
provides general comments on the organization of the atlas. Section~4
gives a basic overview of the principal interstellar lines found in the
FUV, followed by {\S 5} and {\S 6}, which describe the major trends
exhibited by the stellar features as a function of temperature and
luminosity class, respectively. Concluding remarks are given in {\S 7}.

\section{Observations and Data Processing}

The FUV spectra presented here were collected with {\fuse} between 1999
December and 2000 December as part of PI Team programs designed to
investigate the atmospheres of early-type stars, hot gas in the ISM, or
the value of D/H in the local ISM. The {\fuse} instrument consists of four
telescopes and Rowland circle spectrographs and two photon-counting
detectors, which provide redundant coverage of the region between 905 and
1187~{\AA} with a spectral resolving power of $\sim$20,000. Short
wavelength coverage is provided by a pair of telescopes and diffraction
gratings made from silicon carbide (SiC), while efficient coverage of the
region longward of $\sim$1020~{\AA} is provided by a similar pair of
optical elements overcoated with lithium fluoride (LiF). Details of this
instrumentation and its performance during the first year of the mission
have been discussed by \citet{moos00} and \citet{sah00}, respectively. The
brighter targets were observed in histogram (HIST) mode; the remainder were
observed in time-tag (TTAG) mode. All the data presented here were obtained
through the $30\arcsec \times 30\arcsec$ (LWRS) aperture. Exposure times
were typically around 5 ks.

The spectra were processed with the standard {\tt calfuse\/} calibration
pipeline (version 1.8.7). Processing steps included removal of small,
thermally-induced motions of the diffraction gratings; subtraction of a
constant background; correction for thermal and electronic distortions in
the detectors; extraction of a one-dimensional spectrum by summing over
the astigmatic height of the spectrum in the spatial direction; correction 
for detector deadtime; and application of flux and wavelength calibrations.
This version of {\tt calfuse} did not flat field the data or correct for
astigmatism. The end product was a fully calibrated spectrum in the
heliocentric reference frame{\footnote[11]{ Since {\tt calfuse} calibration
pipeline version 1.8.7 did not apply the heliocentric velocity correction
in the proper sense, this step was performed independently.}}. The flux
calibration is accurate to better than 10\%. The wavelength scale has a
precision of about a resolution element ($\sim$15--20 {\kms}), but suffers
from inaccuracy because the zero point for observations through the LWRS
aperture is poorly known.

Since the {\fuse} data are divided between two pairs of channels (SiC1 and 
LiF1; SiC2 and LiF2) which are in turn recorded by two detector segments
(1A and 1B; 2A and 2B), a total of eight independent spectra are generated
for each data set processed by {\tt calfuse}. Unfortunately, it is
difficult to coadd these spectra because of systematic differences in their
spectral resolution and data quality. To avoid these problems, processed
spectra from three channel/detector segment combinations were chosen for
the atlas. Spectra from the LiF2 channel and detector segment A (hereafter
``LiF2A spectra'') were selected to cover the long wavelength region
(1086--1183~{\AA}); SiC1A spectra cover the middle part of the waveband
(1003--1092~{\AA}); and SiC2A spectra illustrate the short wavelength
portion of the {\fuse} waveband (912--1006~{\AA}). These choices represent
the best compromise between the competing requirements of high S/N (i.e.,
large effective area and minimal fixed-pattern noise) and uninterrupted
coverage of the {\fuse} waveband.  The choice of LiF2A spectra to cover the
long-wavelength segment was also determined by the desire to avoid the
anomaly known as ``the worm'' {\citep{sah00}}, which mars LiF1B spectra of
the same region.

As the final step in the processing, the spectra from these three
combinations of channels and detector segments were binned and smoothed
over 20 pixels with a boxcar filter to a nominal spectral resolving 
power of $\sim$8800 (0.12~{\AA}), in order to enhance the visibility
of broad, stellar features. The resultant spectra typically have S/N of
30--35 per 0.12~{\AA} sample in the continua of LiF2A spectra.

\section{Design of the Atlas} 

\subsection{Selection Criteria}

Stars were selected for inclusion in the atlas from the large pool of
Galactic observations based on the degree to which their FUV spectra were
typical of other targets with the same optical spectral classification; the
quality of the observation (high S/N); and the quality of the sight line
(low extinction preferred). For objects with rare spectral classifications,
one or more of these criteria were relaxed in order to improve the coverage
of the temperature- and luminosity-class plane. We occasionally encountered
objects with peculiar or distinctive features compared with other objects
with the same classification. Since Galactic OB stars must be distant
(hence apparently faint) or heavily reddened to satisfy the {\fuse}
instrumental bright limit, many of these objects have not been studied
intensively.  As a result, many of their spectral classifications are based
on old plate material of varying quality, and may not be consistent with
the criteria prevalent today. Unless a recent confirmation of the optical
classification was available, most of these data were rejected.

With these criteria, a total of 45 stars covering spectral types from O2 to
B5 and luminosity classes from dwarfs to supergiants were selected for the
atlas. A few gaps in coverage exist, especially for subgiants and bright
giants which, in any case, are rare among OB stars. Fundamental properties
of the targets are listed in Table~\ref{targets}, where successive columns
record the identity of the object; its J2000 coordinates; its Galactic
coordinates; its spectral classification; the source of the classification;
the apparent visual magnitude; the color excess, $E(\bv)$; the {\fuse}
identification for the data set{\footnote[12]{Interested researchers can
use these identifications to retrieve the pipeline-processed spectra from
the Multi-Mission Archive at Space Telescope (MAST) via
\url{http://archive.stsci.edu/mast.html}. The rebinned and smoothed
versions of the spectra presented in the atlas are available in electronic
form from the MAST Prepared Science Products website
(\url{http://archive.stsci.edu/prep\_ds.html}).}};
and the exposure time in seconds.

\subsection{Organization of the Atlas}

The main sections of the atlas are arranged into five groups according to
the luminosity class of the targets. The spectra for each luminosity class
are presented in three montages, which illustrate the long-, middle-, and
short-wavelength sections of the {\fuse} waveband with LiF2A, SiC1A, and
SiC2A spectra, respectively. Within each montage, the spectra are ordered
by increasing temperature class, i.e., decreasing effective temperature.
Thus, Figures~1 to 3, 4 to 6, 7 to 9, 10 to 12, and 13 to 15 illustrate the
FUV spectra of Galactic dwarfs, subgiants, giants, bright giants, and
supergiants, respectively. The temperature sequences at fixed luminosity
are supplemented by Figures 16 to 18, which show the effects of luminosity
for the spectral type O6-O7.

\notetoeditor{Each Figure (each fXX.ps file) must be at the SAME PHYSICAL SIZE and take ONE JOURNAL PAGE}

\subsection{Description of the Montages}

The FUV spectra plotted in the montages are flux calibrated, with five
spectra per montage. Labels indicate the identity of the object and its
spectral type. The decision to use flux units rather than normalized
spectra was dictated by the enormous difficulty in establishing consistent
continua due to line blending, particularly below 1000~{\AA}. Instead, the
range of fluxes plotted was altered on a star-by-star basis to ensure that
the spectra in successive panels can be compared in a meaningful way. This
approach is feasible because the intrinsic stellar flux distribution is
quite flat over the $\sim$100~{\AA} displayed in each plot, except near
the Lyman limit (where, in any case, the appearance of the spectrum is
often determined by the ISM). However, a consequence of this approach is
that the range of fluxes plotted for a given star is different for the
LiF2A, SiC1A, and SiC2A spectra.

As mentioned earlier, the FUV region of the spectrum is extraordinarily
rich in spectral features arising in the atmospheres and stellar winds of
early-type stars and the many phases of the ISM. Indeed, the typical
density of interstellar lines is so great at shorter wavelengths
(especially due to H$_2$) that stellar continua are strongly depressed,
and line blending is a serious source of confusion. Although detailed
identification of all available transitions is not the main goal of this
atlas, the main features are indicated in the montages in the following
way:
\begin{enumerate}
  \item The rest wavelengths of major stellar features are indicated 
        above the top panel and are also given in Table~\ref{starlines}.
	Components from the same resonance doublet are joined by a line.
  \item The positions of the Lyman series of {\ion{H}{1}} are
        indicated by daggers ($\dagger$) in the upper panel and in
	Table~\ref{starlines}. These are predominantly ISM features
	for spectral types earlier than $\sim$B0.
  \item The positions of H$_2$ lines are indicated by a dense comb
        in the upper panel.  The strength of these features is 
	correlated with the extinction along a given sight line.
	See \S 4 for further discussion.
  \item The positions of other prominent lines in the ISM are indicated
        by a comb in the second panel.  The identities of these lines
	are given in Table~\ref{ismlines}.
  \item The positions of the strongest airglow lines \citep{feld01}
	are indicated by the $\earth$-symbol above the top panel.
	These are also listed in Table~\ref{aglines}.
	Airglow lines appear as narrow emissions, whose strength
	varies from star to star depending on the orbital circumstances
	and the attitude of the spacecraft during the observation.
  \item The positions of known defects in {\fuse} spectra which generally appear 
	to be narrow emission features next to absorptions that often occur
	in the vicinity of airglow lines; see, e.g., the prominent spike near 
	1152~{\AA} in many spectra.
	These locations correspond to regions where the microchannel
	plates have been systematically exposed to more photons during
	the mission. The cumulative effect of this exposure results in fewer
	electrons being liberated by the microchannel plates per input photon
	(a phenomenon known as ``gain sag''), which in turn causes the positions
	of events to be misregistered by the readout circuitry. The strongest
	defects are denoted with ``b'' in Table~\ref{aglines}.
  \item Scattered {\ion{He}{1}} solar emissions from the second order of the 
	gratings, not indicated in the montages, sometimes 
	appear in the \fuse\ spectra similarly to airglow lines. Their 
	strength depends 
	on spacecraft orientation. They are observed at 1044.426, 1074.059, 
	and 1168.668\,\AA, the last being the strongest.
\end{enumerate}

\section{Interstellar Lines} 

From the perspective of stellar spectroscopy, the most
damaging blends come from the strong absorption bands of H$_2$.
The positions of the many lines associated with these bands are
also indicated in the top panel of the montages (see \S 3.3), but this
does not provide an accurate indication of the overall suppression
of flux caused by so many overlapping lines. The devastating effect of
H$_2$ absorption can be gauged {\footnote[13]{In order to distinguish H
and H$_2$ lines from stellar
features, Figures~19 to 21 can be reproduced on transparencies for
use as overlays on the other plots in this atlas. They can also be used to
estimate the column density of H$_2$ along a particular line of sight.}}
from Figures~19
to 21, which show model transmission spectra for a mixture of pure
H and H$_2$ characterized by a single unshifted component with H$_2$
rotational temperature of 300~K, a line width of 5~{\kms},
a fixed ratio of N(H$_2$)/N(\ion{H}{1}) of 0.5, and H$_2$ column densities 
between {$5.0 \times 10^{18}$} and {$1.0 \times 10^{21}$~cm$^{-2}$}.
According to the mean relationship between N(\ion{H}{1}) and 
$E(\bv)$ of \cite{dip94}, this range of {\ion{H}{1}}
column densities corresponds approximately to color excesses of 
{$E(\bv) \le 0.41$}.
With a few exceptions, the color excesses listed in Table~\ref{targets} 
for the stars included in the atlas are between 0.1 and 0.5.
Figures~19 to 21 show that significant bands of flux are removed by H$_2$ 
columns of this magnitude, especially in the shorter wavelength regions.
For rigorous identification of H$_2$ lines see Tables 1 and 2 of 
\citet{mort76}, Table~2 of
\citet{barn00}, and Figures~4 and 5 of \citet{sem99}.
Consequently, for most of the Galactic OB stars presented in this 
atlas, the appearance of the SiC2A spectra (i.e., the spectra
covering the region between 912 and 1006~{\AA}) is determined 
by the ISM and conveys little information about the spectrum of the
star itself. Absorption lines from deuterated hydrogen molecules (HD)
are also visible {\citep[see Table~4 of] []{sem99}}.

Table~\ref{ismlines} provides identifications and rest wavelengths for a 
selection of metal lines that are typically found in the ISM longward of 
Ly~$\delta$
along sight lines to Galactic OB stars. Metallic lines listed in 
Table~\ref{ismlines} are also indicated in the second panel of the montages 
(see \S 3.3). Because {\ion{C}{1}} shows
numerous lines, only the strongest from the ground state with values of 
log($\lambda$f)$\geq$0.4 are considered here.  Although metal lines in the 
ISM are usually much narrower than 
stellar lines, the cumulative effects of blending due to the sheer number of 
transitions and the multi-component nature of the ISM frequently complicate 
the detection and analysis of weak features in the stellar spectrum.
Similarly, the presence of stellar features often complicates the 
determination of the baseline flux (i.e., the local ``continuum'') 
against which an interstellar absorption should be measured.

\section{FUV Spectral Morphology as a Function of Temperature Class}
	 
In this section we describe the trends exhibited by key stellar lines
as a function of the effective temperature for each luminosity class.
Table~\ref{starlines} lists the most prominent stellar features
in the FUV waveband that form the basis for this discussion.
In general, our line identifications are based on \citet{mort91},
\citet{snow76}, \citet{mort77}, \citet{rog85}, \citet{tar97}, and 
the National Institute of Standards and Technology online 
Atomic Spectroscopic Database.\footnote[14]{\url{http://physics.nist.gov/}}
We discuss the behavior of the most important lines in dwarf spectra
in some detail, and thereafter comment primarily on the differences 
exhibited by the same lines in stars of higher luminosity.

\subsection{Dwarfs}

Representative spectra of Galactic O- and B-type dwarfs are shown in 
Figures 1 to 3. Stellar lines from the following species can be identified, and
their behavior tracked.

{\it \ion{H}{1}}. ---
The Lyman series of {\ion{H}{1}} is largely interstellar until $\sim$B1, 
where the broad photospheric component becomes evident.
The series of stellar lines can be followed from Ly~$\beta$ to 
Ly~$\eta$ in the spectrum of HDE\,233622 (B2~V).

{\it \ion{He}{2}}. ---
The appearance of the {\heii~$\lambda$1084.8} line is generally 
compromised by blending with the interstellar lines of 
{\nii~$\lambda\lambda$1084.0, 1084.6, 1085.5, and 1085.7}.
Nevertheless, it is clearly a photospheric feature for all OB dwarfs.
The dramatic strengthening of this feature in spectral types later than 
B0 results from the increasing dominance of {\nii} in the stellar 
atmosphere itself.

{\it \ion{C}{3}}. ---
The {\ciii} resonance line at 977.0~{\AA} is obliterated by blending with 
H$_2$ along the sight lines illustrated in Figure~3. Nevertheless a broad
\ciii~$\lambda$977 absorption is detectable in subtypes O9.5 to B3.
However, the behavior of this transition is studied more easily 
in spectra of OB-type stars in the Magellanic Clouds, which generally 
suffer substantially less reddening and contamination by H$_2$ 
\citep[see][]{mcatlas}.
The {\ciii} multiplet centered at 1175.6~{\AA}, which is due
to transitions between excited levels, is the strongest line
in the LiF2A spectra (Fig.~1).  
It increases slowly in strength with decreasing temperature up
to the $\sim$B0 type. Then the strength is rather constant from B0\,V 
to B3\,V.
Except for HD\,152623 {\citep[O7\,V; see][for a description of this
multiple system]{gar01}}, the line appears to be photospheric, which is
consistent with the expectation that it should be formed preferentially
in regions of higher density. In the case of HD\,152623 the multiplet
exhibits a broad blue-shifted absorption component in 
addition to the strong feature near rest velocity.
This incipient wind profile is more characteristic of evolved stars
(see, e.g., \S 5.3), but has been seen in {\fuse} spectra of 
several other O6--O7 dwarfs.

{\it \ion{C}{4}}. ---
The prominent line blueward of {\ciii~$\lambda$1176} is likely the 
{\civ~$\lambda\lambda$1168.9, 1169.0} doublet, possibly with some contribution 
from the {\niv~$\lambda\lambda$1168.6, 1169.1, 1169.5} multiplet.  
In contrast with the {\ciii} line, the {\civ} feature
exhibits maximum strength for the early O-type dwarfs, weakens
towards the later O stars, and disappears by B0.

{\it \ion{N}{3}}. ---
The {\niii} resonance line at 991.6~{\AA} is largely masked by
interstellar absorption along the sight lines illustrated in Figure~3.
See \citet{mcatlas} for a description of its behavior
in OB-type stars in the Magellanic Clouds.

{\it \ion{N}{4}}. ---
The excited transition of {\niv~$\lambda$955.3} is hidden by 
interstellar absorption along the sight lines illustrated in Figure~3.

{\it \ion{O}{3}}. ---
Three lines, all with similar strength and width are evident in
the spectra of early- to mid- O stars between 1149 and 1154~{\AA}.
These lines strengthen between spectral types O5 and O6, but
disappear abruptly between O8 and O9. These lines are clearly seen
in \fuse\ spectra of late-type (oxygen rich) WC stars.
Consequently we associate them with
{\oiii~$\lambda\lambda$ 1149.6, 1150.9, and 1153.8}.

{\it \ion{O}{6}}. ---
Owing to its unusually high ionization potential (113.896~eV), the
{\ovi~$\lambda\lambda$1032, 1038} doublet is the most intriguing 
feature in FUV spectra of OB-type stars.
It typically exhibits a single P~Cygni wind profile for the early O stars,
with one broad absorption trough blended with the saturated interstellar
Ly~$\beta$ line, and a redshifted emission peak near 1040~{\AA}.
In later O-type spectra, two distinct blueshifted absorptions are often
visible, though any emission that might be present is effectively
removed through absorption by the saturated interstellar 
{\cii~$\lambda\lambda$1036.3, 1037.0} doublet.
For a detailed discussion concerning the origin of this exotic
species, see {\citet{mac93}}.

{\it \ion{Si}{3}}. ---
Many transitions due to {\siiii} are apparent in spectra cooler than $\sim$O9.
The most prominent are the {\siiii}~$\lambda\lambda$1108.5 1110.8, 1113.2
triplet, which continues to strengthen until at least B2.
Other prominent features in the B0 -- B2 range, particularly the
broad absorption between 1140 and 1145~{\AA} and the blended features
between 1155 and $\sim$1173~{\AA}, are also due to {\siiii}.

{\it \ion{Si}{4}}. ---
The {\siiv~$\lambda\lambda$1122.5, 1128.3} doublet is photospheric
in the dwarf spectra illustrated in Figure~1.
The blue component is blended with IS lines, while the
red component is blended with the {\pv~$\lambda$1128.0} stellar feature.
While the {\pv} line weakens with decreasing temperature, the
{\siiv} lines strengthen, so that they dominate the blend at $\sim$B0
and are still strong at B2.
Thus, the overall appearance of the {\pv+\siiv} $\lambda$1128 blend
remains nearly constant along the temperature sequence.
Another strong {\siiv} line at 1066.6~{\AA} strengthens similarly
for late O and early B-type stars.

{\it \ion{P}{4}}. ---
The {\piv}~$\lambda$950.6 resonance line cannot be disentangled from
blending with interstellar features along the sight lines illustrated
in Figure~3.

{\it \ion{P}{5}}. ---
The {\pv~$\lambda\lambda$1118.0, 1128.0} resonance doublet is an important
diagnostic because it is expected to be the dominant stage of ionization
for most O stars, but is seldom saturated because of the low cosmic
abundance of phosphorus.
The lines weaken substantially around B0, and vanish by $\sim$B2.
The behavior of the red component is complicated by a blend with
{\siiv~$\lambda$1128.3}, which shows the opposite temperature dependence.

{\it \ion{S}{4}}. ---
The {\siv}~$\lambda\lambda$1073.0, 1073.5 lines (and also $\lambda$1062.7,
though this component is frequently blended with a strong
H$_2$ feature) are largely photospheric in dwarf spectra.
Although rather weak at early O types, these lines quickly increase
in strength for late O stars.
Other {\siv} lines are also prominent features of the FUV spectra of
late O and early B spectral types.
In particular, the strong feature at $\lambda$1099.4 noticed by
\citet{sta75} and \citet{wal96} has now been shown to
be due to {\siv} {\citep{wer01}}.
Weaker {\siv} lines might also blend with the {\siiii} lines at 
1108.5, 1110.8, and 1113.2~{\AA}.
Yet another {\siv} line is located at 1138.2~{\AA}, which overlaps
with the extended blend of {\siiii} lines between 1140 and 1145~{\AA}.

{\it \ion{S}{6}}. ---
The visibility of the {\svi}~$\lambda\lambda$933.3, 944.5 resonance lines
in early O-type stars is often compromised by blending with H$_2$ lines
and the confluence of IS Lyman lines of {\ion{H}{1}} \citep[see e.g.][]{mcatlas}.
The extinction along the Galactic sight lines illustrated in Figure~3 
precludes the detection of this important diagnostic.

\subsection{Subgiants}

FUV spectra of a small sample of subgiants are illustrated in Figures 4 to 6.
This luminosity class is poorly defined for O-type stars, and representatives
earlier than B0 are rare.
Consequently, our sample is limited to temperature classes between O9 and B2.
The stellar lines prominent in dwarf spectra  also dominate the
appearance of subdwarf spectra, and their behavior as a function of
temperature follows the same trends.

\subsection{Giants}

Figures 7 to 9 show the FUV spectra of giants with temperature classes
between O5 and B3.  
The lines from several species exhibit different morphologies or
behavior compared with the dwarfs.

{\it \ion{He}{2}}. ---
The {\nii} component of the blend begins to dominate the appearance
of this feature in giant spectra by $\sim$O9.5.

{\it \ion{C}{3}}. ---
P~Cygni profiles can be seen in the {\ciii~$\lambda$977.0} resonance
line for spectral types earlier than O9, though (as with the dwarfs)
visibility is determined mostly by the degree of interstellar contamination.
In contrast to its appearance in dwarf and subdwarf spectra, 
the excited {\ciii~$\lambda$1176} multiplet now exhibits broad, blueshifted 
absorption in addition to a narrower, undisplaced photospheric component
for giants hotter than O9.5.
Weak, redshifted emission is also present, though its visibility is
compromised because it falls near the edge of the detector segment
and is distorted by blends with weak stellar lines. The
{\ciii}~$\lambda$1176 photospheric absorption strengthens
systematically in the spectra of giants between O9.5 and at least B3.

{\it \ion{C}{4}}. ---
The line at 1169~{\AA} attributed to {\civ} and/or {\niv} is still
observed but is also blended with the absorption part of the {\ciii}
wind profile for spectral types earlier than $\sim$B0.

{\it \ion{O}{6}}. ---
The {\ovi~$\lambda\lambda$1032, 1038} resonance doublet shows
a strong P~Cygni profile in the spectra of HD\,93843 (O5~III(f)var) with both
shifted emission and absorption features. The wind profile shows
a nice sequence of decreasing intensity from O5 to O7.
The blue-shifted absorption trough extends beyond the interstellar
Ly~{$\beta$} line for spectral types earlier than O7.
Although the appearance of the {\ovi} wind feature can be quite
different for the O9 stars observed with \fuse\ 
(HD\,116852 is possibly misclassified), 
there is an overall trend for the feature 
to decrease toward cooler temperatures.
It weakens substantially near $\sim$B0, as shown in Figure~8 by the 
increasingly symmetric appearance of the Ly~{$\beta$} interstellar 
(and, after $\sim$B1, stellar) line. 
Blueshifted absorptions can be identified with {\ovi} in spectra
as late as B1, though their identification is frequently
difficult {\citep[see,~e.g.,][]{leh01}}.

{\it \ion{P}{5}}. ---
In contrast to its appearance in spectra of less luminous stars,
the {\pv~$\lambda\lambda$1118, 1128} resonance doublet exhibits
P~Cygni profiles in the spectra of giants earlier than O7.
The blue component of the doublet weakens substantially 
between about B1 and B2, but the red component appears similar throughout
owing to the increasing dominance of the blend with the 
{\siiv~$\lambda$1128} line.

{\it \ion{S}{4}}. ---
The {\siv~$\lambda\lambda$1063, 1073} resonance lines reach
maximum strength near B0 in giant spectra.
Weak P~Cygni profiles might be present at mid-O spectral types,
but otherwise these lines appear to be photospheric.
The {\siv}~$\lambda$1099.4 feature strengthens until O9.5 and
disappears by B2.

{\it \ion{S}{6}}. ---
Despite severe interstellar contamination, the presence of broad, 
blue-shifted absorption can be detected in the 
{\svi~$\lambda\lambda$933, 944} resonance doublet for spectral types 
up to O9.

\subsection{Bright Giants}

Only a few early-type bright giants have been observed with {\fuse}.
The FUV spectra of stars between O7 and B1.5 are presented in Figures
10 through 12.
The appearance of the key diagnostic lines is intermediate between
their appearance in giant and supergiant spectra, and their behavior as
a function of temperature is also similar.
The sudden switch in morphology of the {\ciii~$\lambda$1176} between
a wind profile at O9.7 and a strong photospheric absorption at
B0.5 is particularly striking.

\subsection{Supergiants}

A selection of FUV spectra of Galactic OB supergiants is presented in 
Figures~13 through 15.  
In sharp contrast to the less luminous objects, stellar wind
features dominate the appearance of these spectra.
The major morphological trends are as follows.
 
{\it \ion{He}{2}}. --
Although badly blended with interstellar lines of {\nii}, the
{\heii~$\lambda$1084.8} line is purely photospheric in character, 
and does not appear to change strength substantially between O2 and O9.
The increase in strength for spectral types later than $\sim$O9.5
is due to the predominance of stellar {\nii}, which contributes negligibly
to the blend for hotter stars.

{\it \ion{C}{3}}. ---
Broad P~Cygni absorption troughs from the {\ciii~$\lambda$977}
resonance line can be detected in spectra from O6 to O9.7,
despite severe contamination with interstellar lines that removes
a large central portion of the profile.
The absorption trough generally engulfs the interstellar Ly~{$\gamma$} 
line of {\ion{H}{1}}.
The {\ciii~$\lambda$1176} feature behaves similarly, though its
astonishing morphological changes are much more clearly visible.
It exists as a weak photospheric line between O2 and O4, but
suddenly emerges as a well developed, nearly saturated P~Cygni
profile by O6.
Since the {\ciii~$\lambda$1176} line is due to transitions between
excited levels, the strength of these P~Cygni profiles implies that
the wind is very dense.
The wind profile vanishes abruptly between B0 and B1, to be replaced
by a very strong, unshifted absorption feature.
Note that the behavior of \ciii~$\lambda$1176 in supergiant spectra
is identical to that of \siiv~$\lambda\lambda$1394, 1403, which has
a similar ionization potential {\citep{wal96}}.

{\it \ion{C}{4}}. ---
The {\civ} and/or {\niv} line near 1169~{\AA} increases in strength 
from O2 to O4, before being lost in the absorption troughs of the 
strong {\ciii~$\lambda$1176} P~Cygni profiles between O6 and O9.  
It is present at O9.7, but weakens substantially in the early B stars.

{\it \ion{N}{3}}. ---
Traces of a P~Cygni profile can be seen in the {\niii~$\lambda$991}
resonance line around O9-O9.7, though it is strongly
contaminated by absorption from interstellar H$_2$ for the sight lines
illustrated in Figure~15.

{\it \ion{O}{6}}. ---
The {\ovi~$\lambda\lambda$1032, 1038} resonance doublet appears as a strong 
P~Cygni profile for supergiants from O2 to B0.
Excess absorption attributable to {\ovi} is still visible at B1.
The blue edge of the absorption trough usually extends beyond the blue wing 
of the interstellar Ly~{$\beta$} line, though the extent of the wind 
absorption decreases with decreasing temperature.
\citet{tar97} analyzed FUV spectra of HD\,93129A, the hottest supergiant 
in our sample, obtained with BEFS during the {\it ORFEUS-SPAS I} mission.
They measured a wind terminal velocity of 3200~{\kms} from the {\ovi} 
P~Cygni profile, which is entirely consistent with the {\fuse} observation 
shown in Figure~14.

{\it \ion{Si}{4}}. ---
As with the other luminosity classes, the {\siiv~$\lambda\lambda$1122, 1128}
lines begin to strengthen at $\sim$B0, the line at 1128\AA\ being dominated
by {\pv} in hotter stars.

{\it \ion{P}{5}}. ---
The {\pv~$\lambda\lambda$1118, 1128} resonance doublet exhibits
P~Cygni profiles for all supergiants between O2 and O9.7.
A blueshifted absorption is seen at O2. Suddenly a strong wind
profile appears at O4,
which then become less pronounced with decreasing temperature.
The {\siiv~$\lambda$1128.3} line begins to dominate the red
component of the doublet at $\sim$B0, which also marks the
appearance of an unshifted absorption at 1118.552~{\AA} that might
be due to {\piv} {\citep{rog85}}.

{\it \ion{S}{4}}. ---
The {\siv~$\lambda\lambda$1062, 1073} lines exhibit dramatic
morphological changes as a function of temperature similar to
that observed for the {\ciii~$\lambda$1176} line.
The {\siv} lines (i.e., the {\siv~$\lambda$1073} line, since the
blue component is badly blended with interstellar H$_2$ lines) are
not present at O2, and weakly present as blueshifted absorption at O4.
However, by O6 they are strong, fully developed P~Cygni profiles,
which persist until $\sim$B0, when they again abruptly alter their
appearance to become strong but unshifted, rotationally broadened
absorption lines.
The absorption lines remain strong until at least B2.
The {\siv~$\lambda$1099.4} feature also appears near O4, strengthens
to a maximum near B0, and persists until at least B2.

{\it \ion{S}{6}}. ---
Although the flux distributions for spectral types earlier than
$\sim$O7 suggest that P~Cygni profiles might be present in 
the lines of the {\svi~$\lambda\lambda$933, 944} resonance doublet,
interstellar contamination is too severe along the sight lines presented 
in Figure~15 to be definite.

\section{FUV Spectral Morphology as a Function of Luminosity Class}

Figures~16, 17, and 18 show LiF2A, SiC1A, and SiC2A spectra, respectively,
for dwarf to supergiant luminosity classes of temperature type O6-O7.
These figures serve as the basis for discussing the most important
luminosity effects visible in FUV spectra of OB stars.
Since higher luminosity increases the mass-loss rate and hence density 
of a radiatively driven stellar wind (all other things being equal),
luminosity effects are particularly evident in the strength and morphology of 
stellar wind profiles.

{\it \ion{C}{3}}. ---
\citet{snow76} and \citet{wal96} have previously discussed the
extraordinary changes in the morphology of {\ciii~$\lambda$1175.6}
with increased luminosity, which accompany the equally dramatic
changes as a function of temperature described in \S 5.
For the mid-O spectra shown in Figure~16, the line exhibits a smooth transition
from a rotationally broadened, unshifted photospheric profile (dwarfs),
to a weak P~Cygni profile (giants), to a strong, nearly saturated
P~Cygni profile (supergiants) as the luminosity increases.
This sensitivity to wind density is attributable to the origin of
this multiplet in transitions between excited states of {\ciii}, which
require dense environments to be populated.
However, for spectral types later than $\sim$B1, the strength
and morphology of the {\ciii~$\lambda$1175} line loses its sensitivity
to luminosity altogether, evidently because it is no longer formed in
the stellar wind.
The broad photospheric absorption line in B stars appears with more or less
the same strength for all luminosity classes. For the hotter stars from
O2 to O4, where the wind profile is not developed yet, the faint photospheric
absorption tends to be deeper for main sequence stars.
The behavior of the {\ciii~$\lambda$1176} line with luminosity class is
analogous to the behavior of the {\siiv~$\lambda\lambda$1394, 1403} doublet,
and all other lines of similar ionization potential.

{\it \ion{C}{4}}. ---
For the O2 to O4 stars, the {\civ} and/or {\niv} line near 1169\AA\ mirrors the
behavior of the {\ciii~$\lambda$1175} line.
 
{\it \ion{O}{6}}. ---
The appearance of the {\ovi~$\lambda\lambda$1032, 1038} 
doublet does not depend strongly on luminosity for O-type stars,
probably because of saturation effects as for {\nv~$\lambda\lambda$1239, 1243} 
and {\civ~$\lambda\lambda$1548, 1551}.
In Figure~17, all luminosity classes exhibit very strong, well-formed
P~Cygni profiles. However, for cooler, B-type spectra, a trend with 
luminosity does exist:
although the {\ovi} wind profile remains strong for B0 supergiants,
it is much weaker (and correspondingly harder to detect) in the 
spectra of B0 dwarfs.

{\it \ion{P}{5}}. ---
The {\pv~$\lambda\lambda$1118, 1128} resonance lines show a similar, 
though less dramatic, luminosity effect to the {\ciii~$\lambda$1175.6} feature.
In mid-O spectra, where they achieve their greatest strength, they
are pure absorption lines for dwarfs, weak P~Cygni profiles for
giants, and stronger P~Cygni profiles for supergiants.
See Figure~16.

{\it \ion{S}{4}}. ---
\citet{wal96} first drew attention to the strong luminosity effect exhibited
by the {\siv~$\lambda\lambda$1063, 1073} lines.
As shown in Figure~17, these lines change at mid-O spectral types from pure 
absorptions in dwarf spectra, to weak blueshifted wind profiles in giant spectra, 
to fully formed P~Cygni profiles in supergiant spectra.
However, in the early B spectra the {\siv} lines are practically
independent of luminosity class. These trends closely follow those exhibited by 
{\ciii~$\lambda$1175.6} and {\pv~$\lambda\lambda$1118, 1128},
which is perhaps not surprising given the close relationship between
their temperature variations (\S 5).

\section{Concluding Remarks} 

The primary result emerging from this atlas is a refined understanding of 
the trends in the strength and morphology of the wide variety of spectral 
lines found in FUV spectra of OB-type stars.
It is remarkable that these FUV features -- the most prominent of which 
arise in the stellar wind -- vary so smoothly according to spectral type 
and luminosity class, which are defined by photospheric lines in the 
optical region of the spectrum. These smooth variations were already
observed from IUE data by Walborn, Nichols-Bohlin, \& Panek (1985).
An obstacle to the detection of these correlations is the accuracy of the 
optical spectral classifications.
Some of the classifications for the Galactic targets accessible to {\fuse} 
are quite dated, and may not have been classified according to the same 
criteria currently used for brighter standards.
To correct this situation, we are currently collecting new optical spectra 
for some of these stars, so that their published spectral types can 
be verified.

This atlas also confirms previous results {\citep[e.g.][]{snow76, wal96}}
concerning the sensitivity of various FUV lines to physical conditions in
the stellar atmosphere. In particular, the strength and morphology of the
{\ciii~$\lambda$1175.6} line is shown to be a strong function of both
effective temperature and luminosity. Similar behavior is demonstrated for
the {\siv~$\lambda\lambda$1063, 1073} and {\pv~$\lambda\lambda$1118, 1128}
lines.  The {\ion{O}{6}~$\lambda\lambda$1032, 1038} wind profiles, which
persist to early B-type stars, exhibit few variations as a function of
luminosity.

Finally, this spectral atlas illustrates directly the treasure-trove of
astrophysical information that lies encoded in FUV spectra. The broad
wavelength coverage, sensitivity, and good spectral resolution of {\fuse}
provide many opportunities to study the atmospheres of early-type stars
and the intervening ISM in fundamentally new ways. As one application of
this new capability, we are currently adding {\fuse} spectra of Galactic
OB-type stars to the databases used by population synthesis codes
{\citep[e.g., Dionne \& Robert, in preparation;][]{ star99}} in order to
use the FUV line sensitivities noted above to refine studies of young
stellar populations in distant galaxies.

\acknowledgments

AP and CR acknowledge financial support from the Natural Sciences and
Engineering Research Council of Canada and the Universit{\'e} Laval.
LB acknowledges support from NASA LTSA grant NAG5-9219 (NRA99-01-LTSA-029).
US participants in the {\fuse} PI-team acknowledge NASA for
support through contract NAS5-32985.

\clearpage


\begin{figure}
\figurenum{1}
\plotone{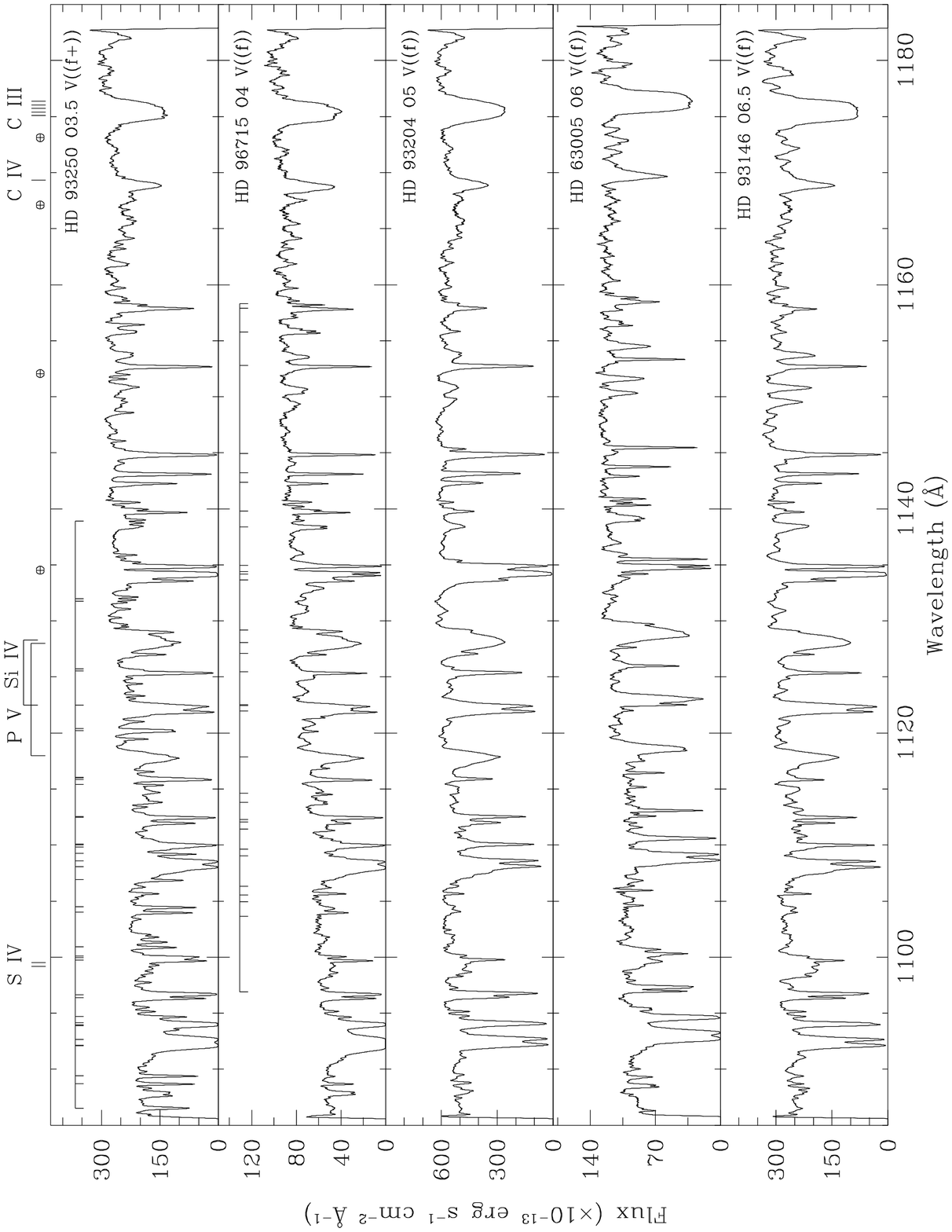}
\caption {\fuse\ spectra of dwarf stars with spectral types from (a) O3.5 to O6.5,
(b) O7 to B0, and (c) B0.5 to B3 between 1085 and 1185\AA. Strong stellar 
lines are identified at the top of each figure. Interstellar lines are
indicated by a dense comb in the upper panel (H$_2$), and second panel
(prominent lines from other species). Airglow lines are identified by $\oplus$.}
\end{figure}

\clearpage

\begin{figure}
\figurenum{1b}
\plotone{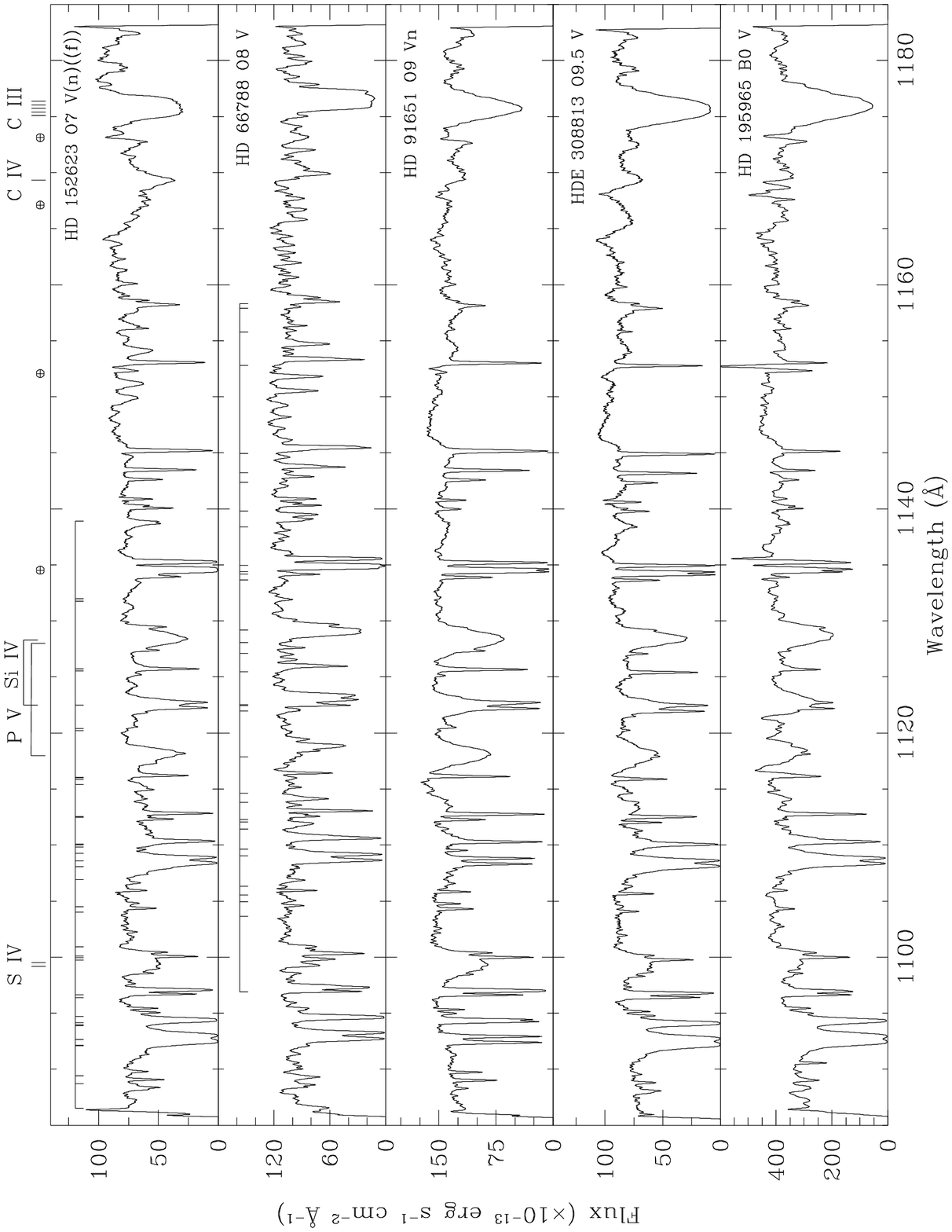}
\end{figure}

\clearpage

\begin{figure}
\figurenum{1c}
\plotone{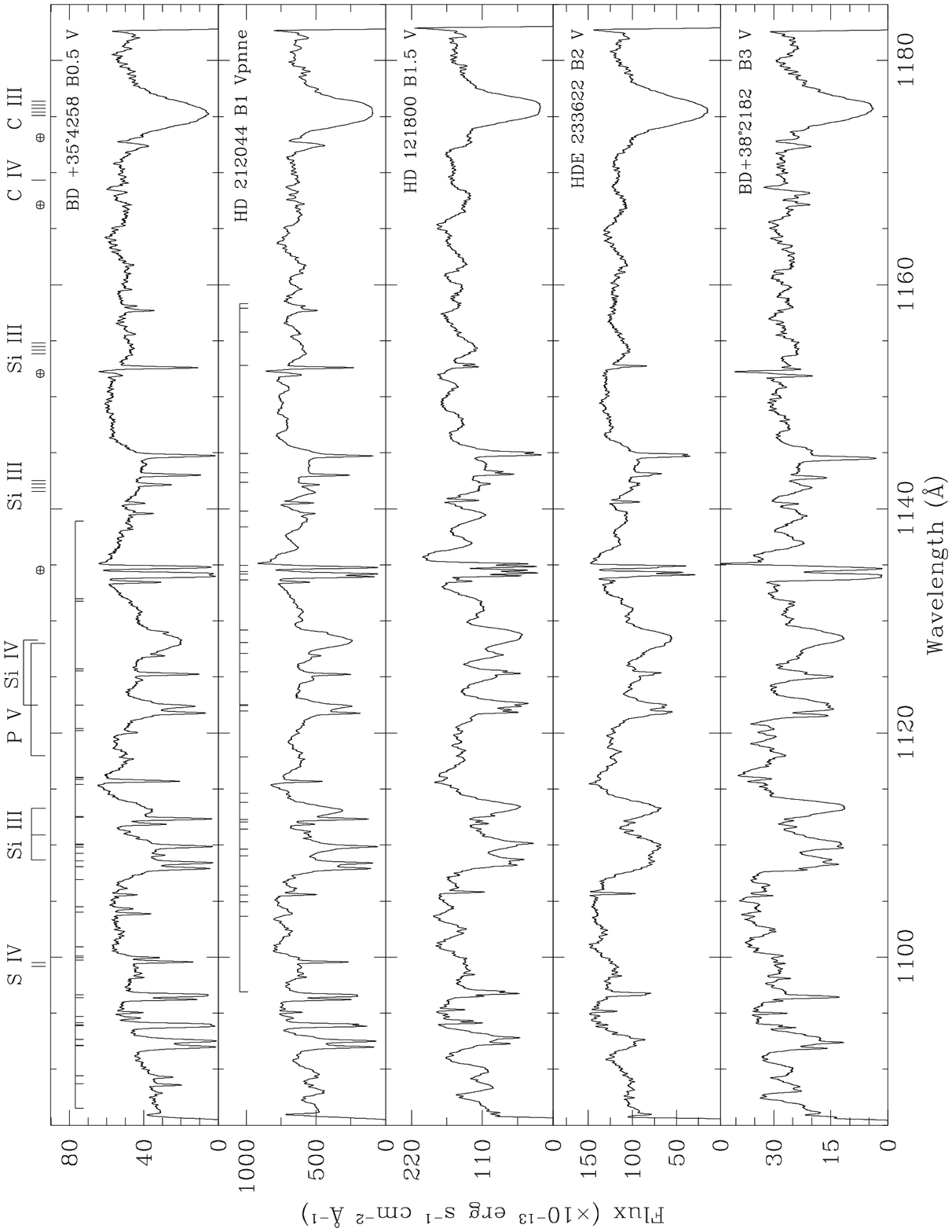}
\end{figure}

\clearpage

\begin{figure}
\figurenum{2}
\plotone{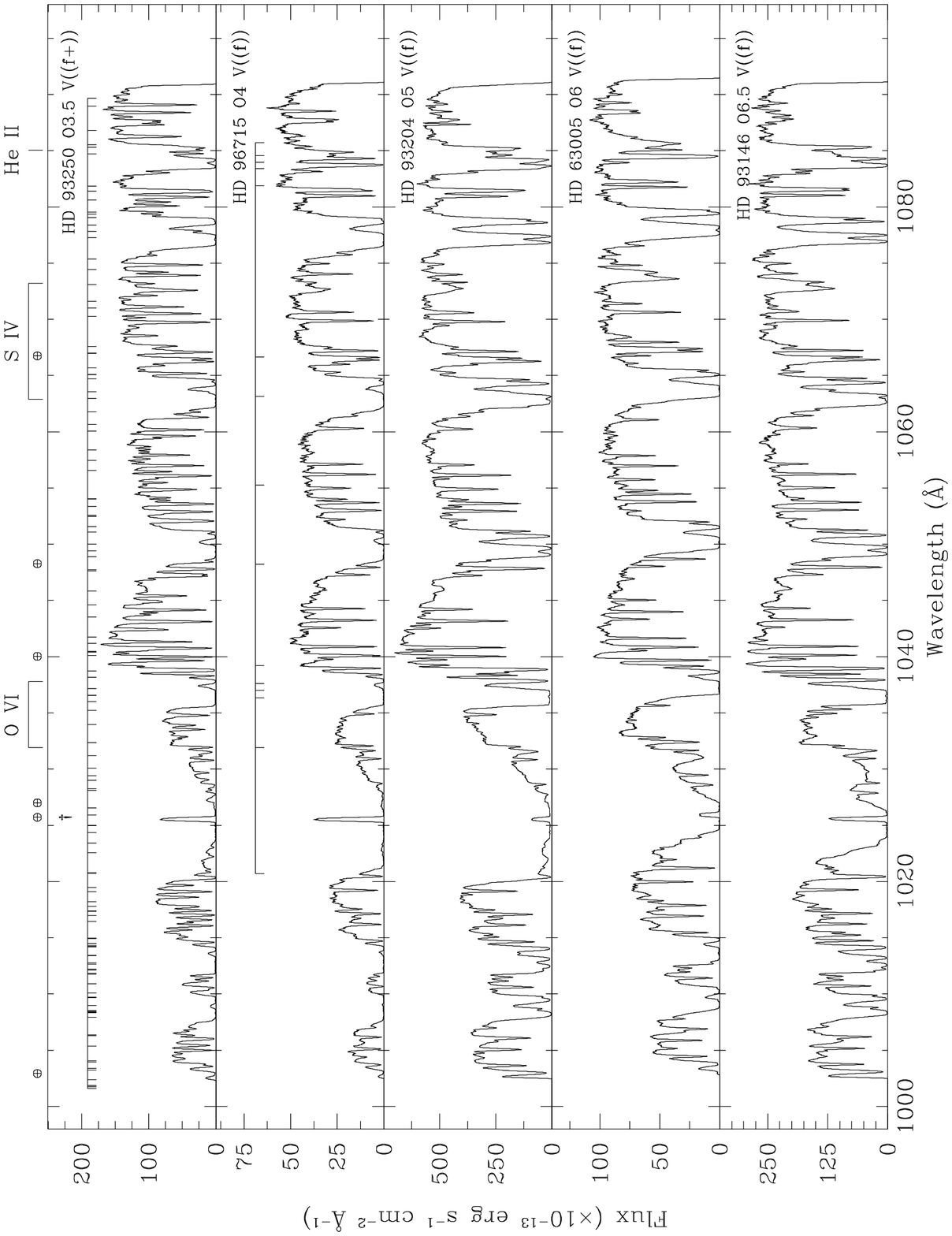}
\caption {\fuse\ spectra of dwarf stars with spectral types from (a) O3.5 to O6.5,
(b) O7 to B0, and (c) B0.5 to B3 between 998 and 1098\AA. See Figure~1a for a description of the labels.}
\end{figure}

\clearpage

\begin{figure}
\figurenum{2b}
\plotone{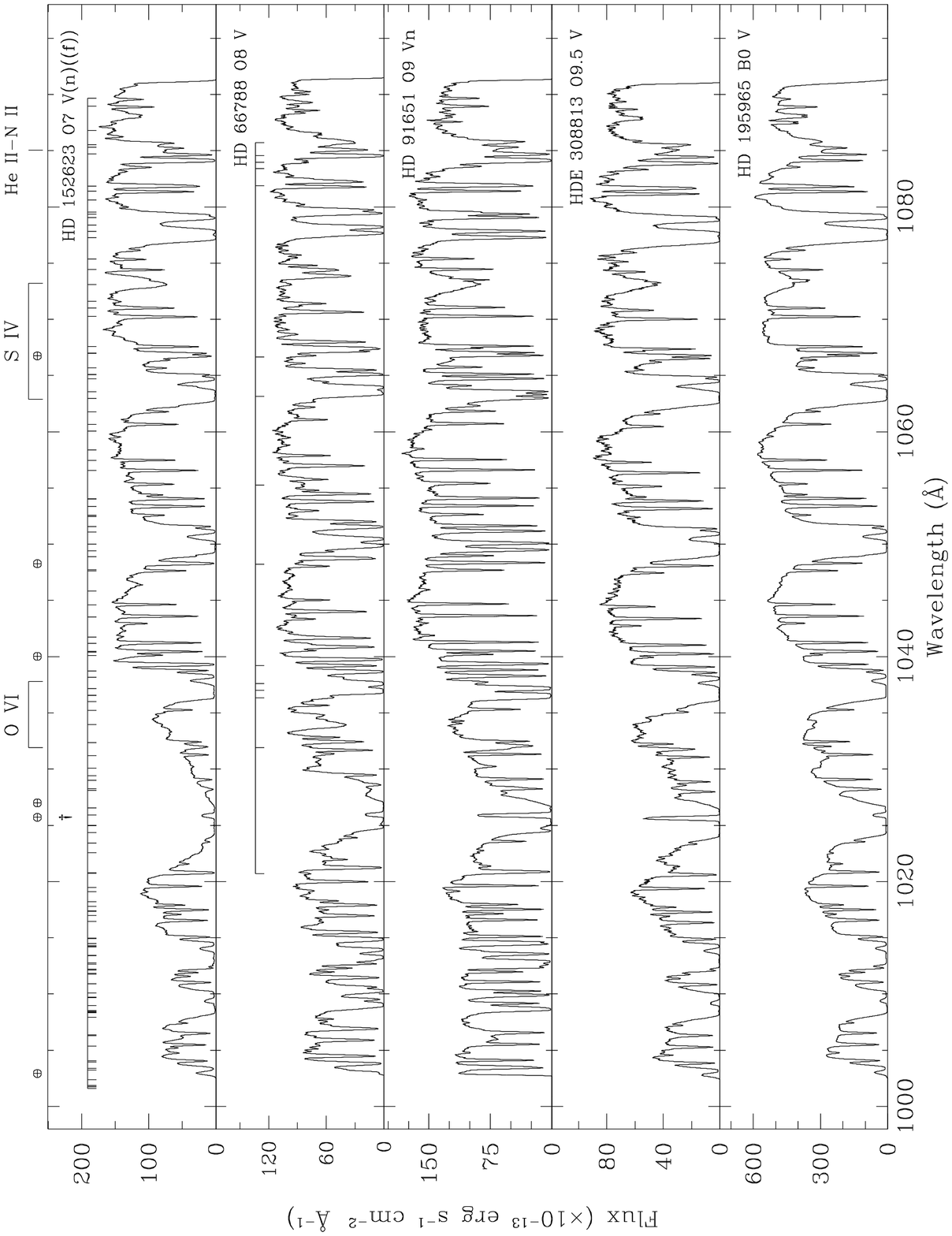}
\end{figure}

\clearpage

\begin{figure}
\figurenum{2c}
\plotone{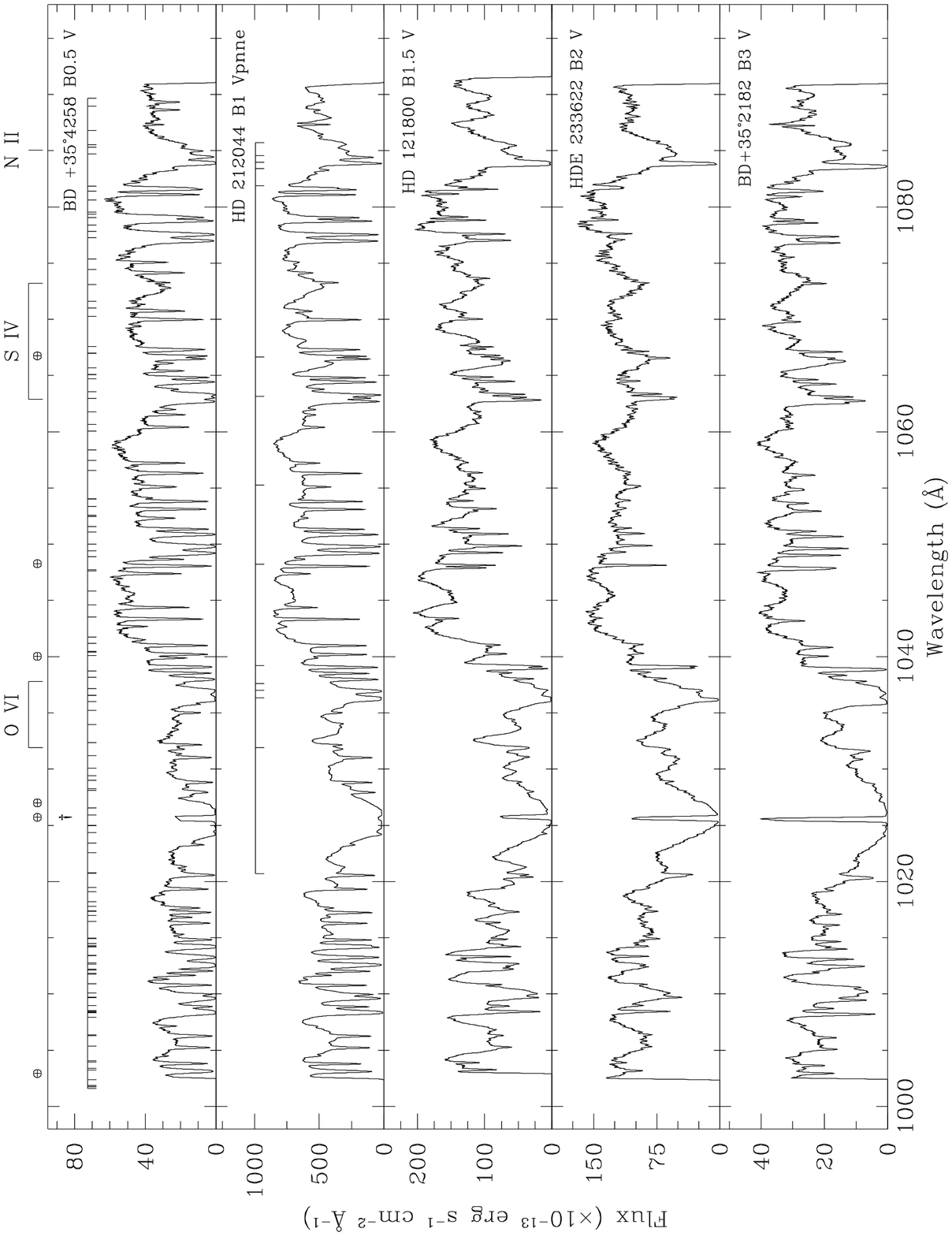}
\end{figure}

\clearpage

\begin{figure}
\figurenum{3}
\plotone{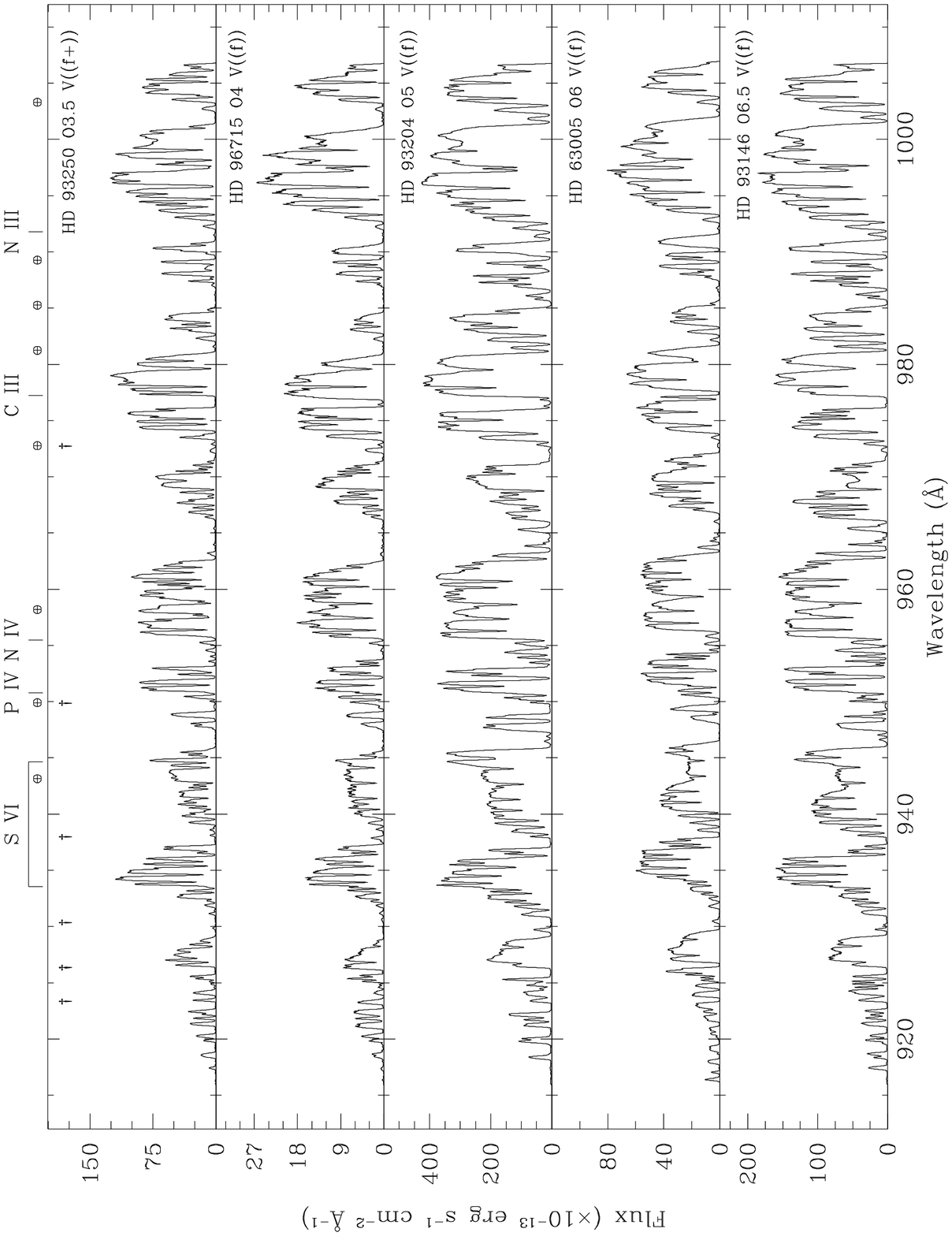}
\caption {\fuse\ spectra of dwarf stars with spectral types from (a) O3.5 to O6.5,
(b) O7 to B0, and (c) B0.5 to B3 between 912 and 1012\AA. See Figure~1a for a description of the labels.}
\end{figure}

\clearpage

\begin{figure}
\figurenum{3b}
\plotone{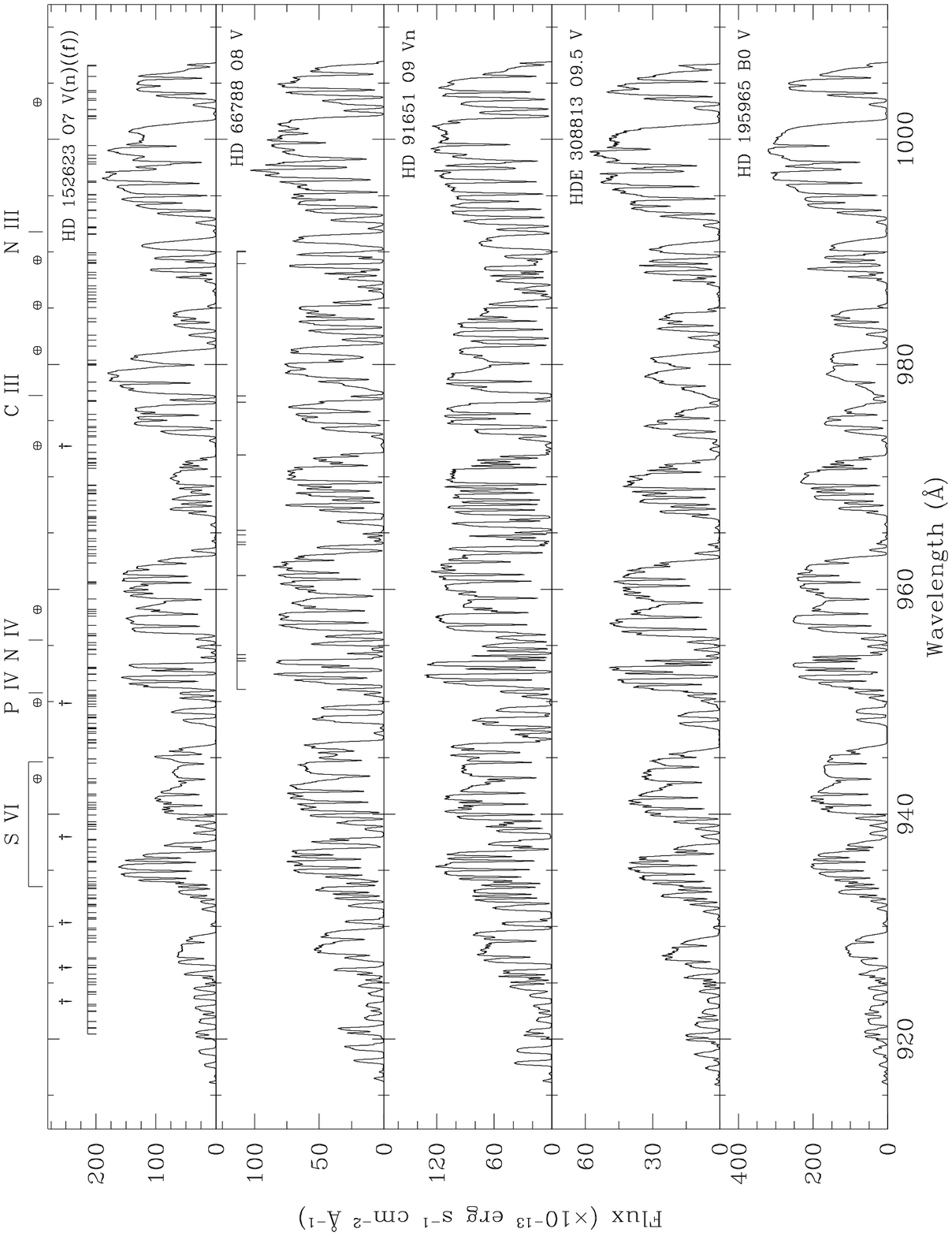}
\end{figure}

\clearpage

\begin{figure}
\figurenum{3c}
\plotone{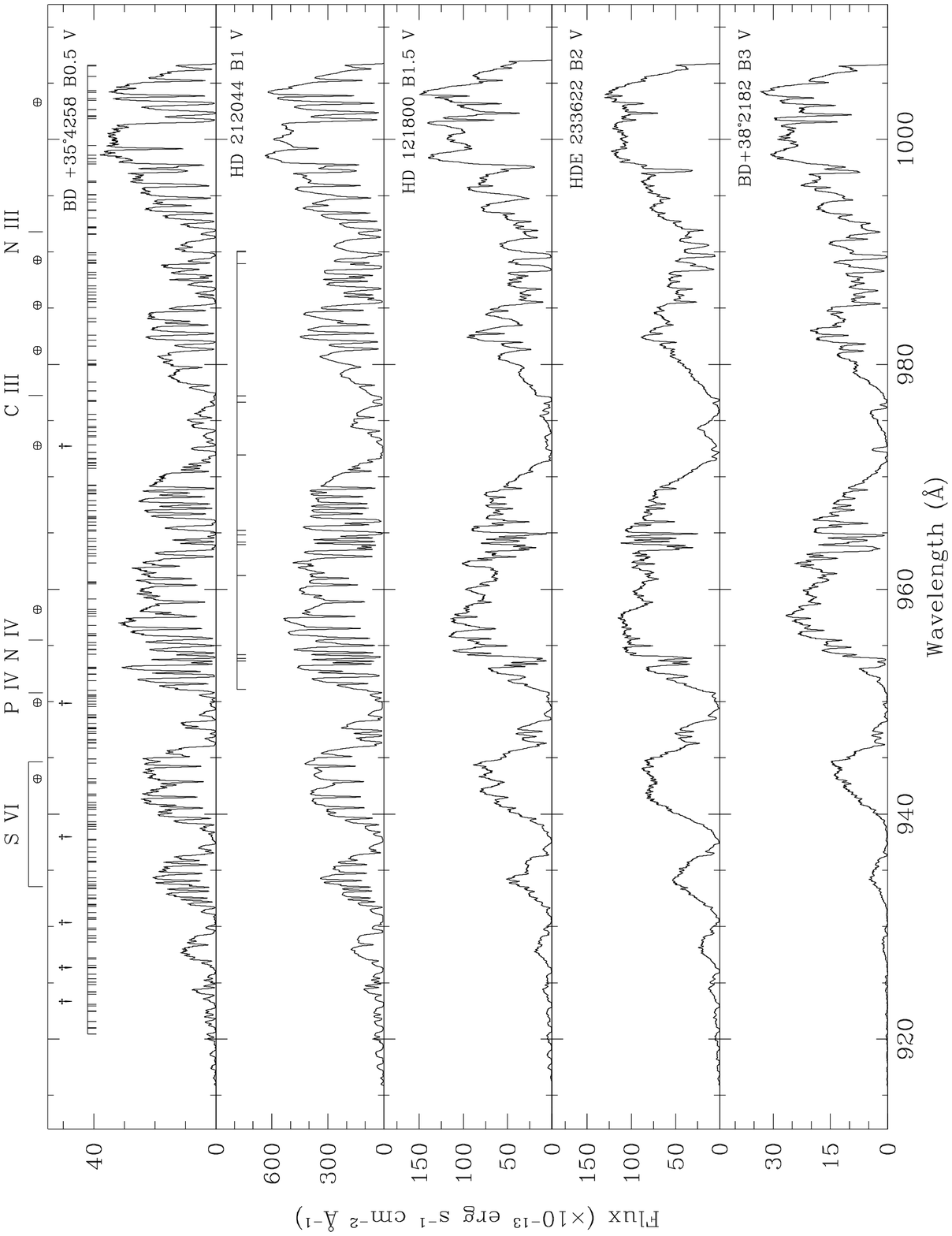}
\end{figure}

\clearpage

\begin{figure}
\figurenum{4}
\plotone{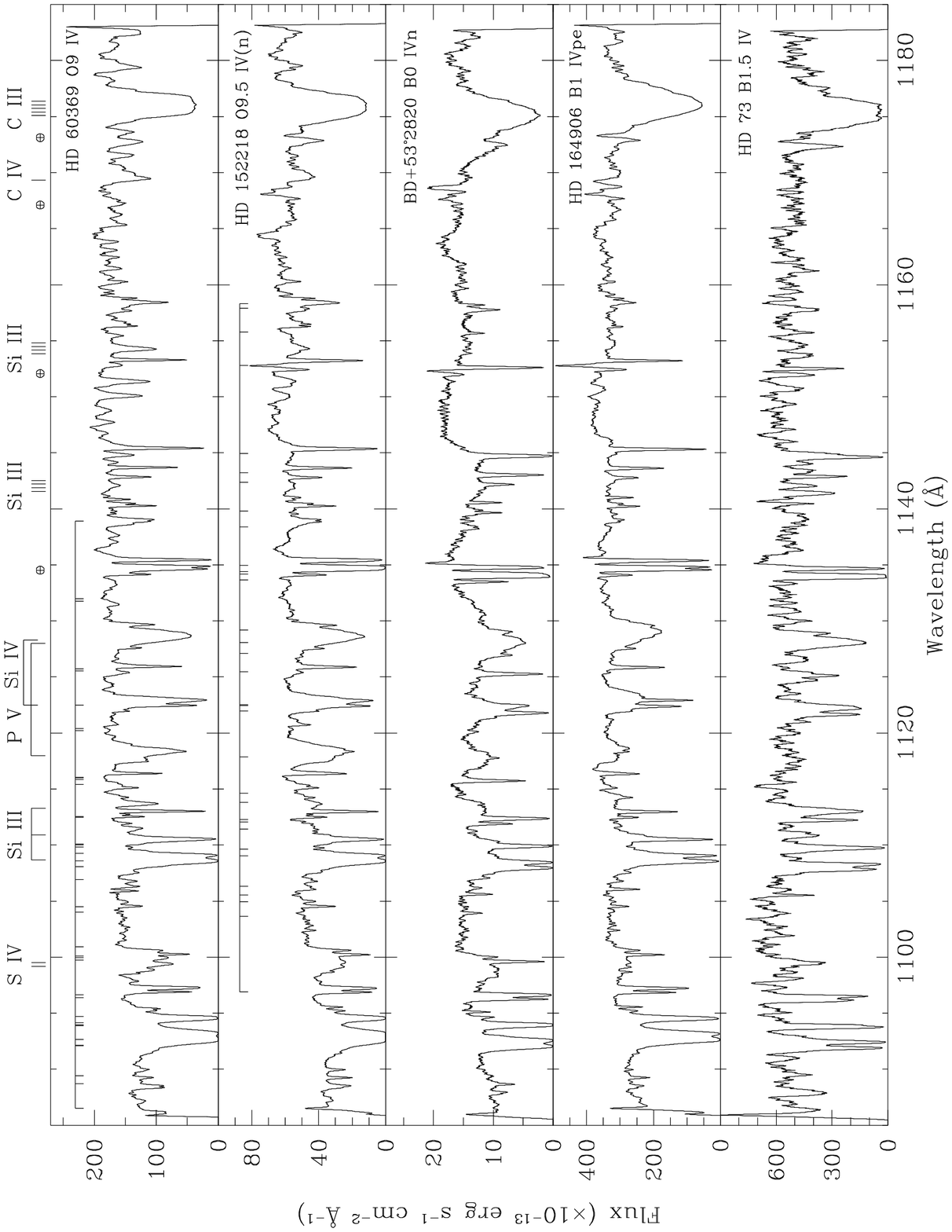}
\caption {\fuse\ spectra of subgiant stars with spectral types from O9 to B1.5
between 1085 and 1185\AA. See Figure~1a for a description of the labels.}
\end{figure}

\clearpage

\begin{figure}
\figurenum{5}
\plotone{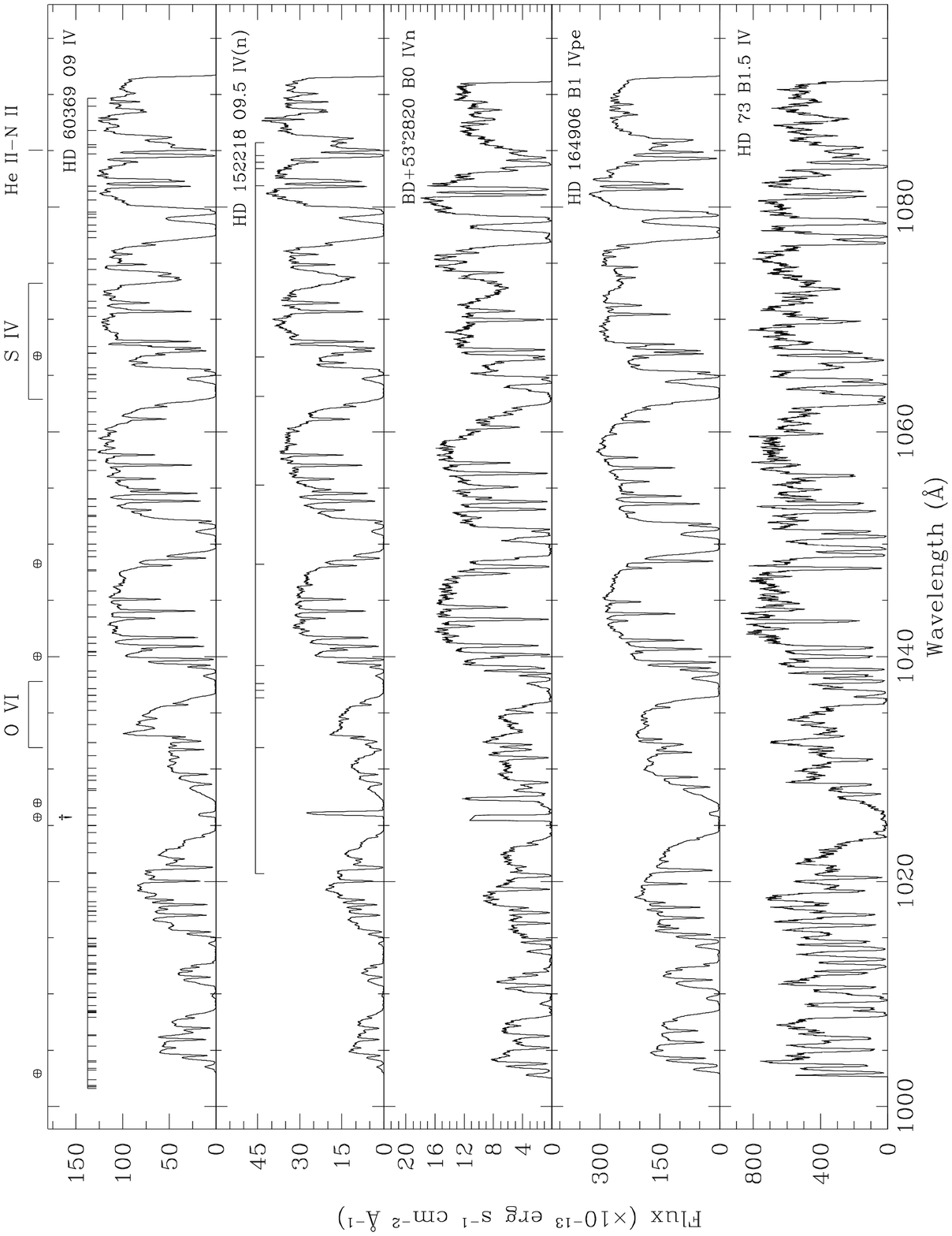}
\caption {\fuse\ spectra of subgiant stars with spectral types from O9 to B1.5
between 998 and 1098\AA. See Figure~1a for a description of the labels.}
\end{figure}

\clearpage

\begin{figure}
\figurenum{6}
\plotone{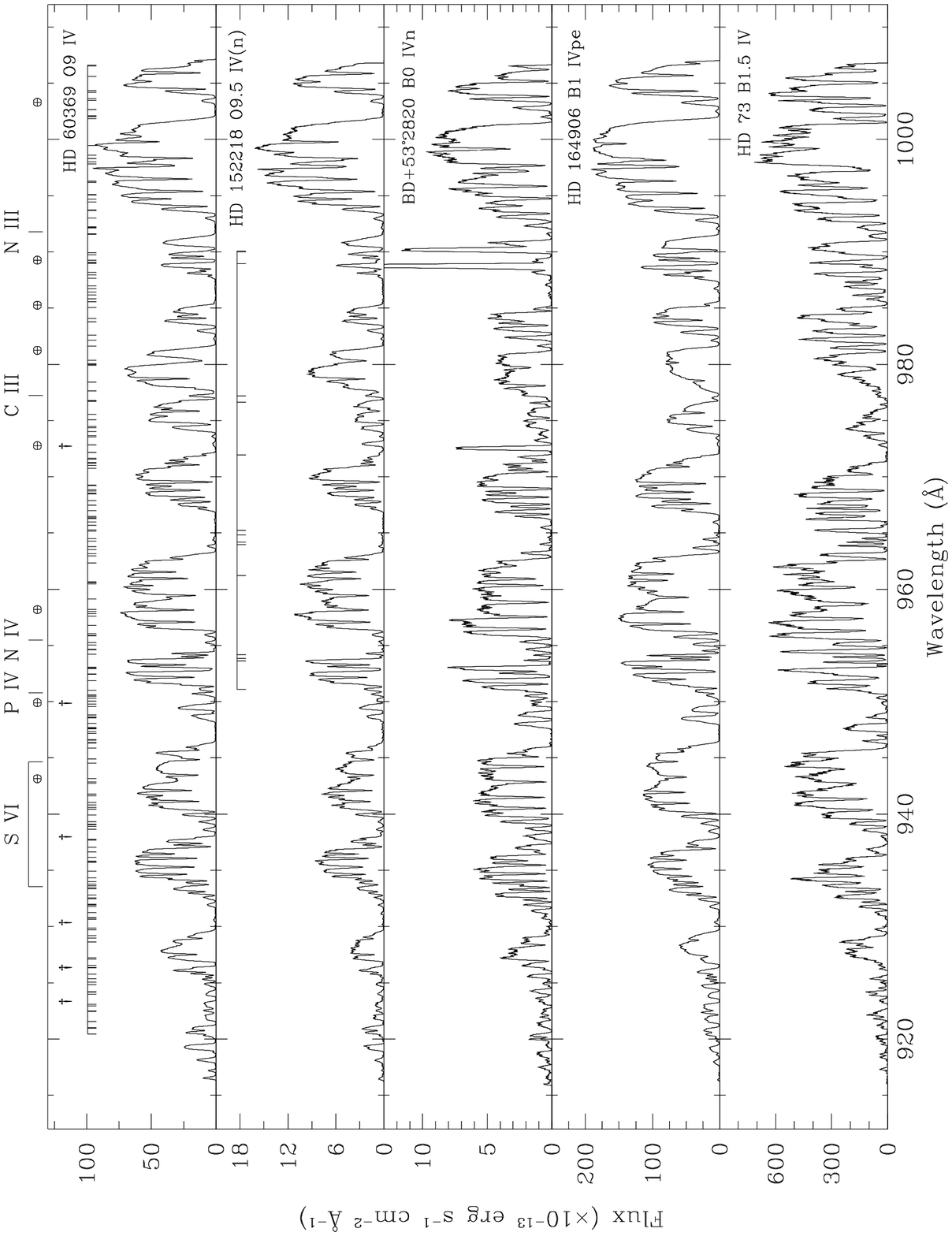}
\caption {\fuse\ spectra of subgiant stars with spectral types from O9 to B1.5
between 912 and 1012\AA. See Figure~1a for a description of the labels.}
\end{figure}

\clearpage

\begin{figure}
\figurenum{7}
\plotone{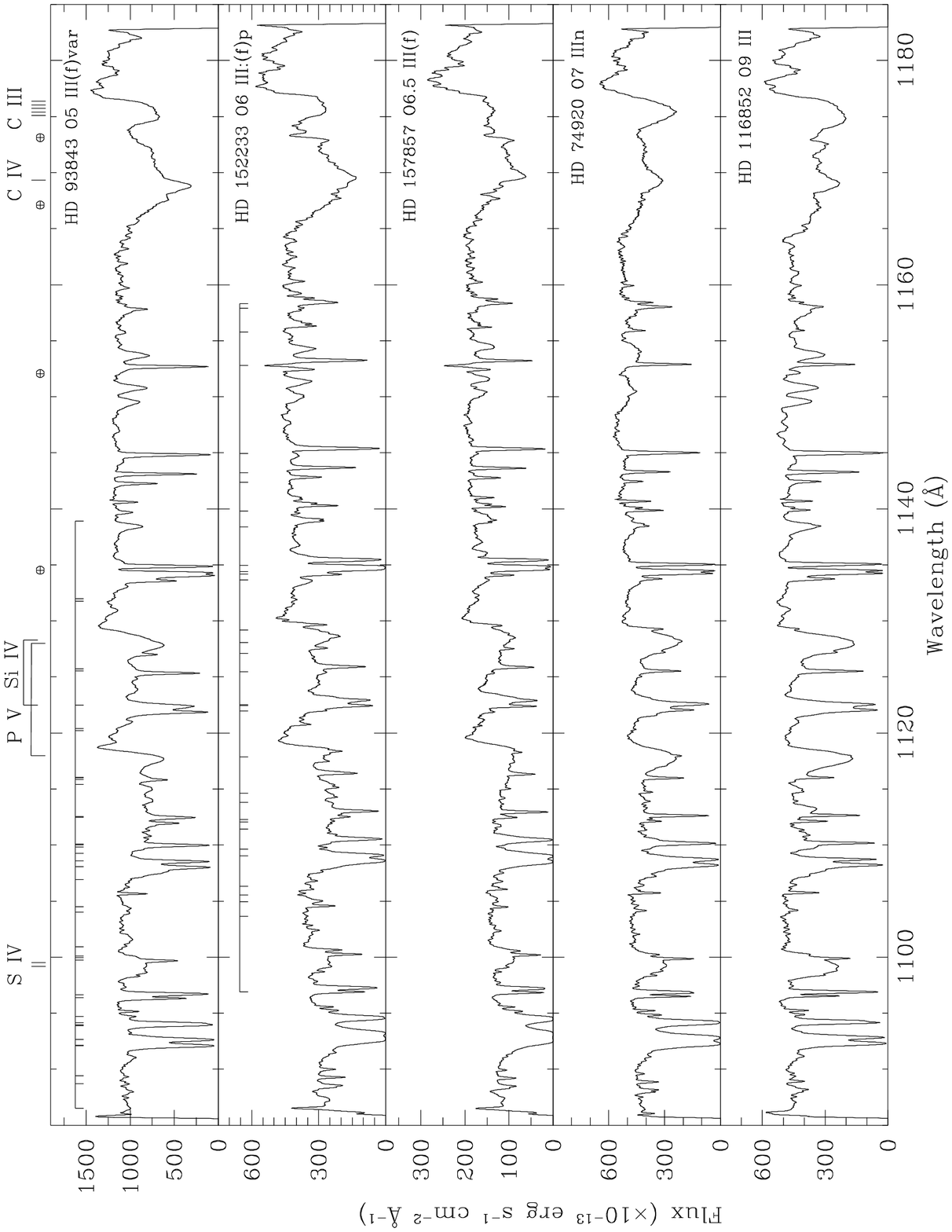}
\caption {\fuse\ spectra of giant stars with spectral types from (a) O5 to O9, and
(b) O9.5 to B3 between 1085 and 1185\AA. See Figure~1a for a description of the labels.}
\end{figure}

\clearpage

\begin{figure}
\figurenum{7b}
\plotone{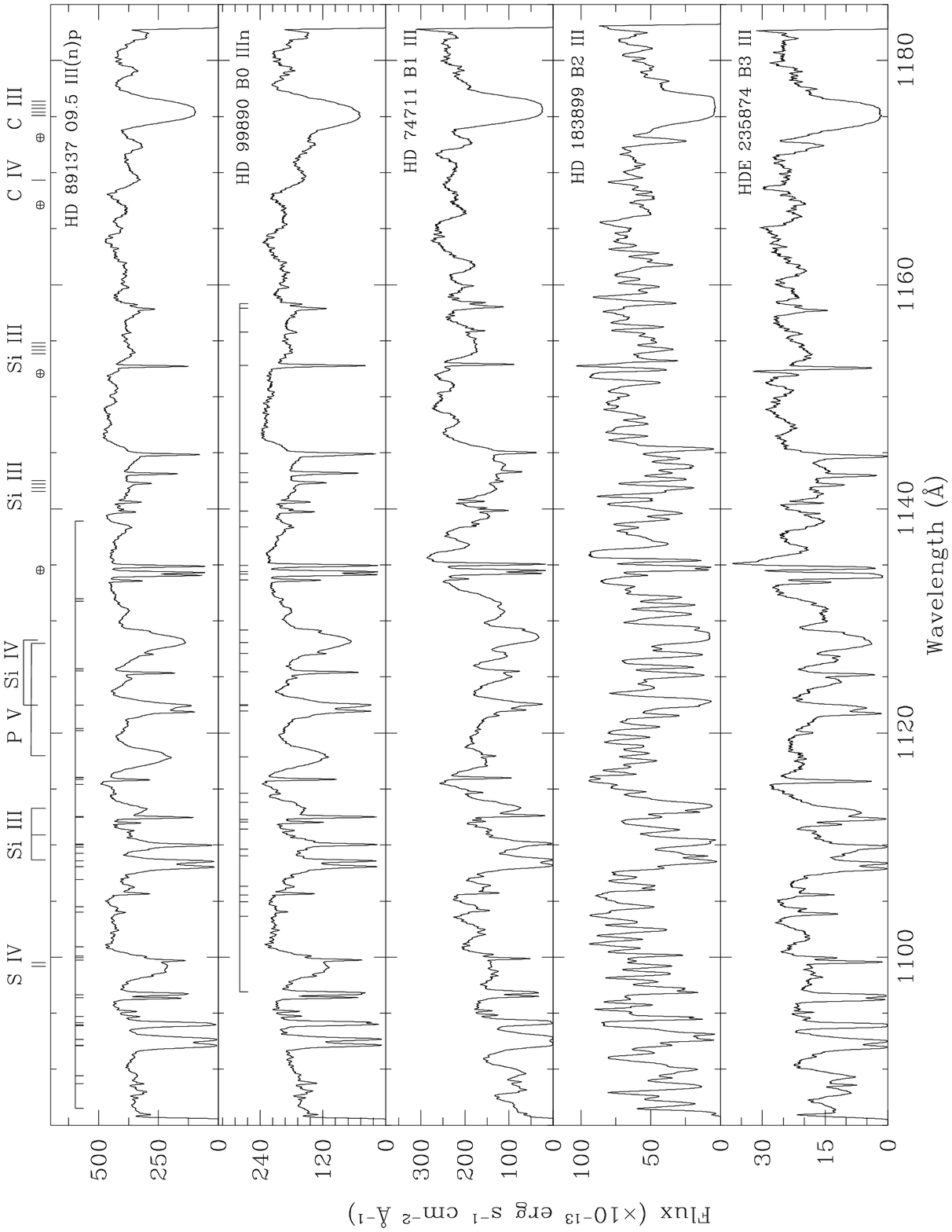}
\end{figure}

\clearpage

\begin{figure}
\figurenum{8}
\plotone{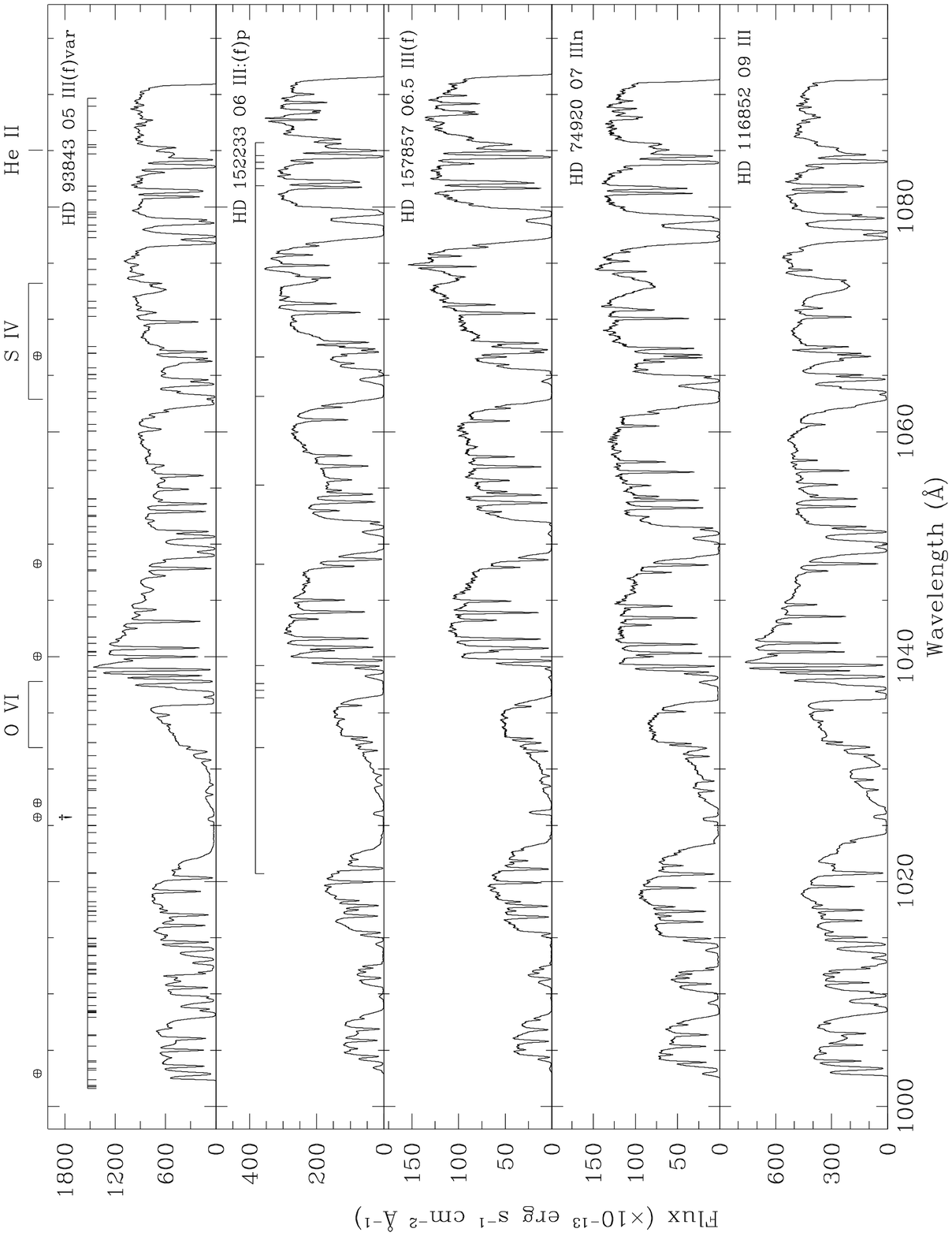}
\caption {\fuse\ spectra of giant stars with spectral types from (a) O5 to O9, and
(b) O9.5 to B3 between 998 and 1098\AA. See Figure~1a for a description of the labels.}
\end{figure}

\clearpage

\begin{figure}
\figurenum{8b}
\plotone{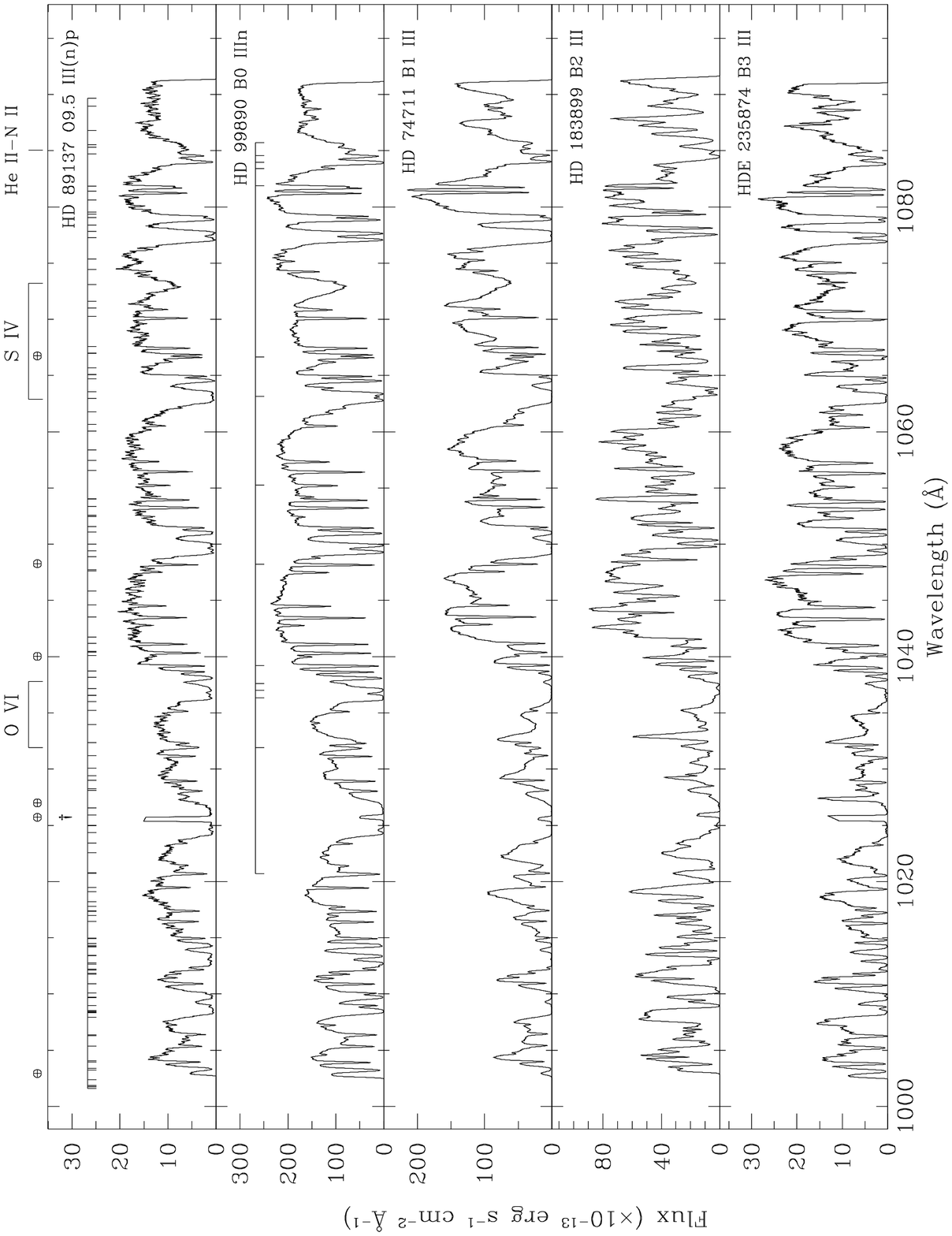}
\end{figure}

\clearpage

\begin{figure}
\figurenum{9}
\plotone{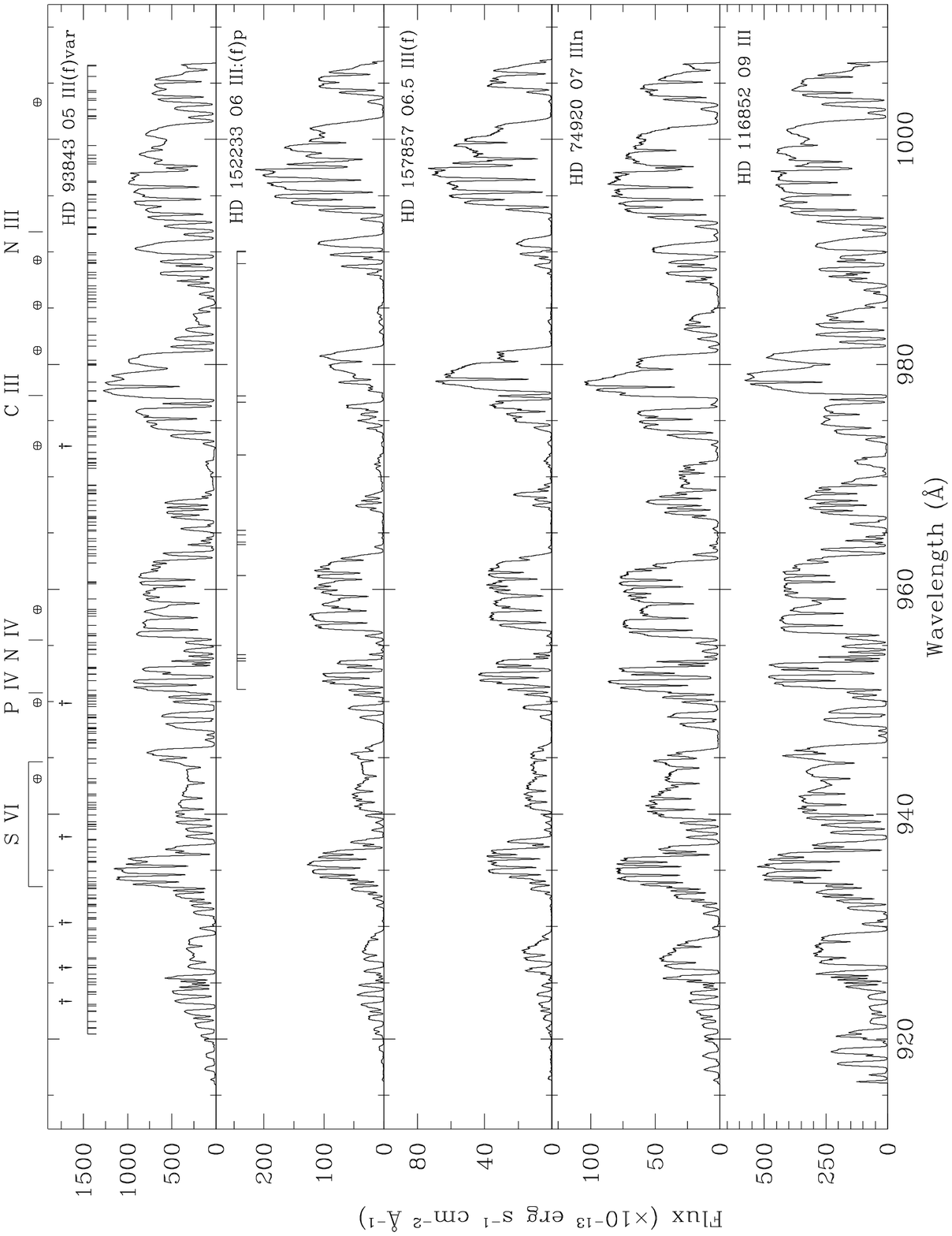}
\caption {\fuse\ spectra of giant stars with spectral types from (a) O5 to O9, and
(b) O9.5 to B3 between 912 and 1012\AA. See Figure~1a for a description of the labels.}
\end{figure}

\clearpage

\begin{figure}
\figurenum{9b}
\plotone{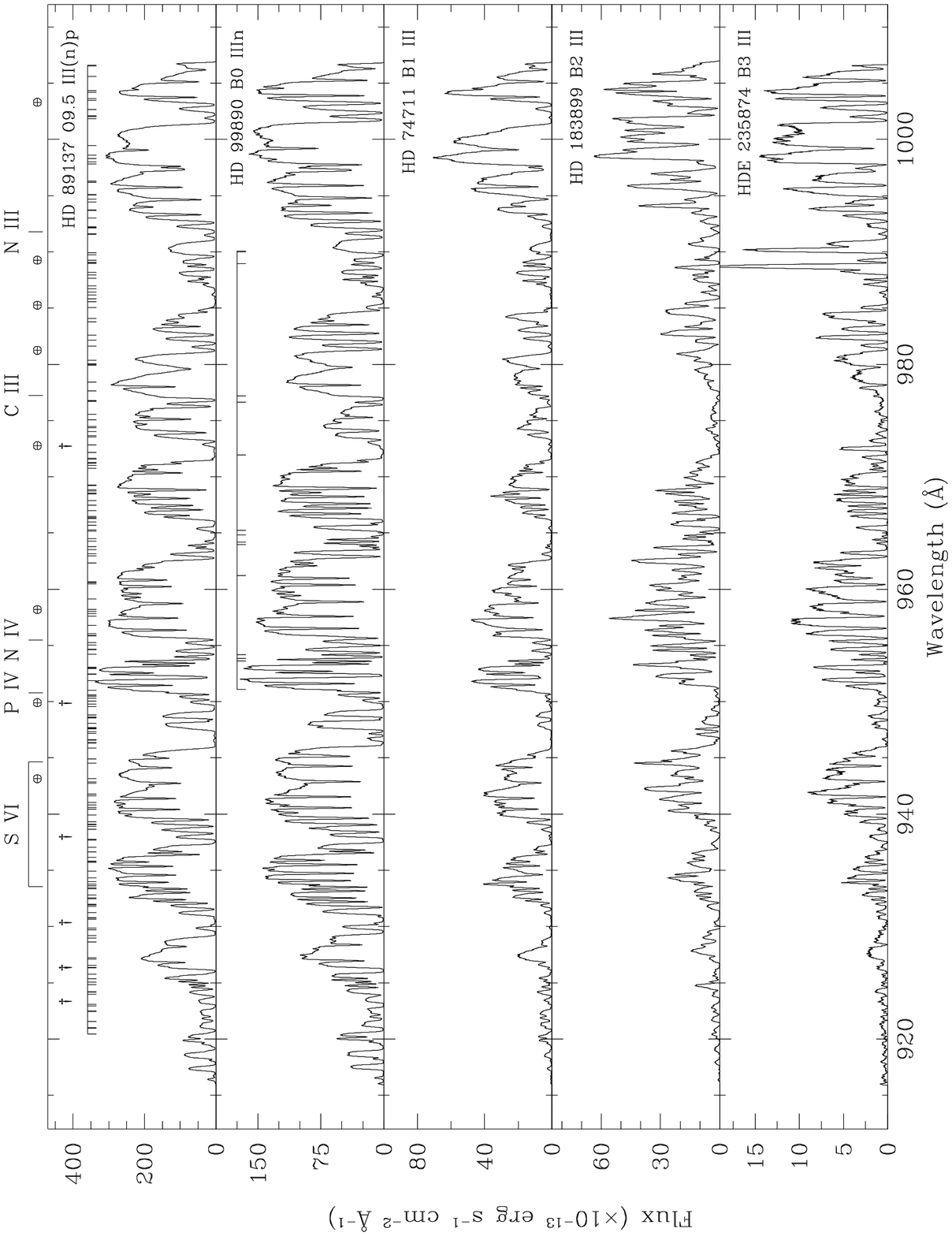}
\end{figure}

\clearpage


\begin{figure}
\figurenum{10}
\plotone{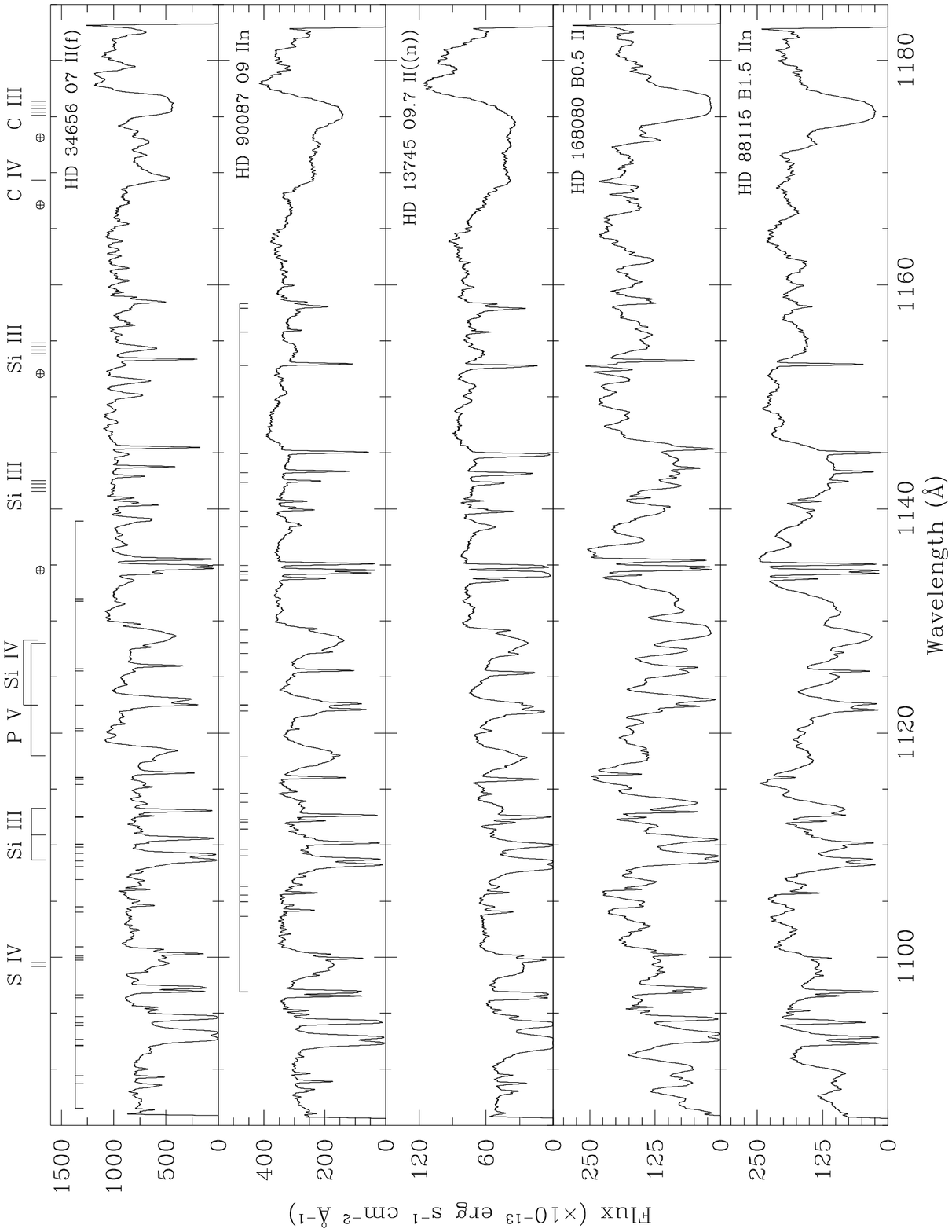}
\caption {\fuse\ spectra of bright giant stars with spectral types from O7 to B1.5
between 1085 and 1185\AA. See Figure~1a for a description of the labels.}
\end{figure}

\clearpage

\begin{figure}
\figurenum{11}
\plotone{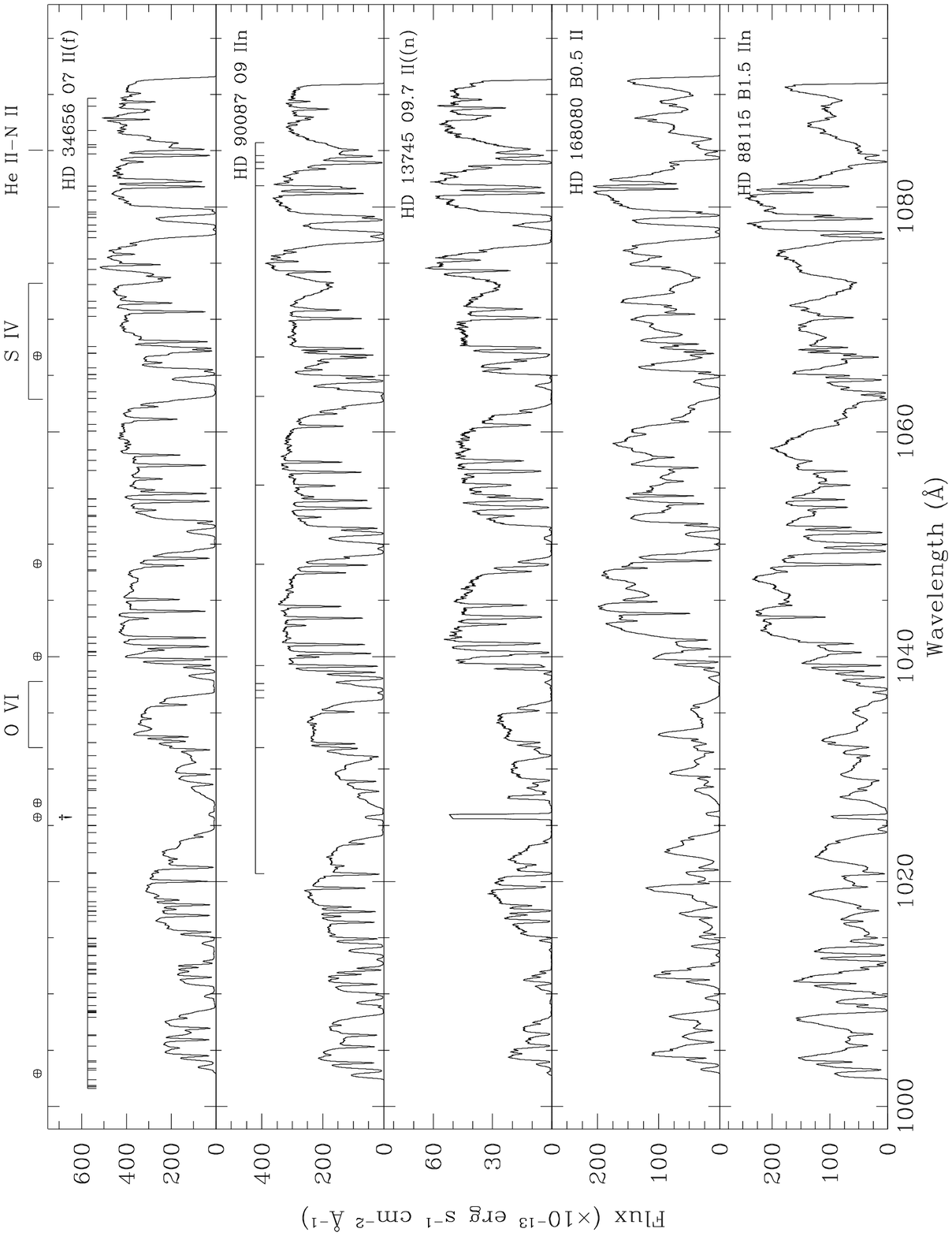}
\caption {\fuse\ spectra of bright giant stars with spectral types from O7 to B1.5
between 998 and 1098\AA. See Figure~1a for a description of the labels.}
\end{figure}

\clearpage

\begin{figure}
\figurenum{12}
\plotone{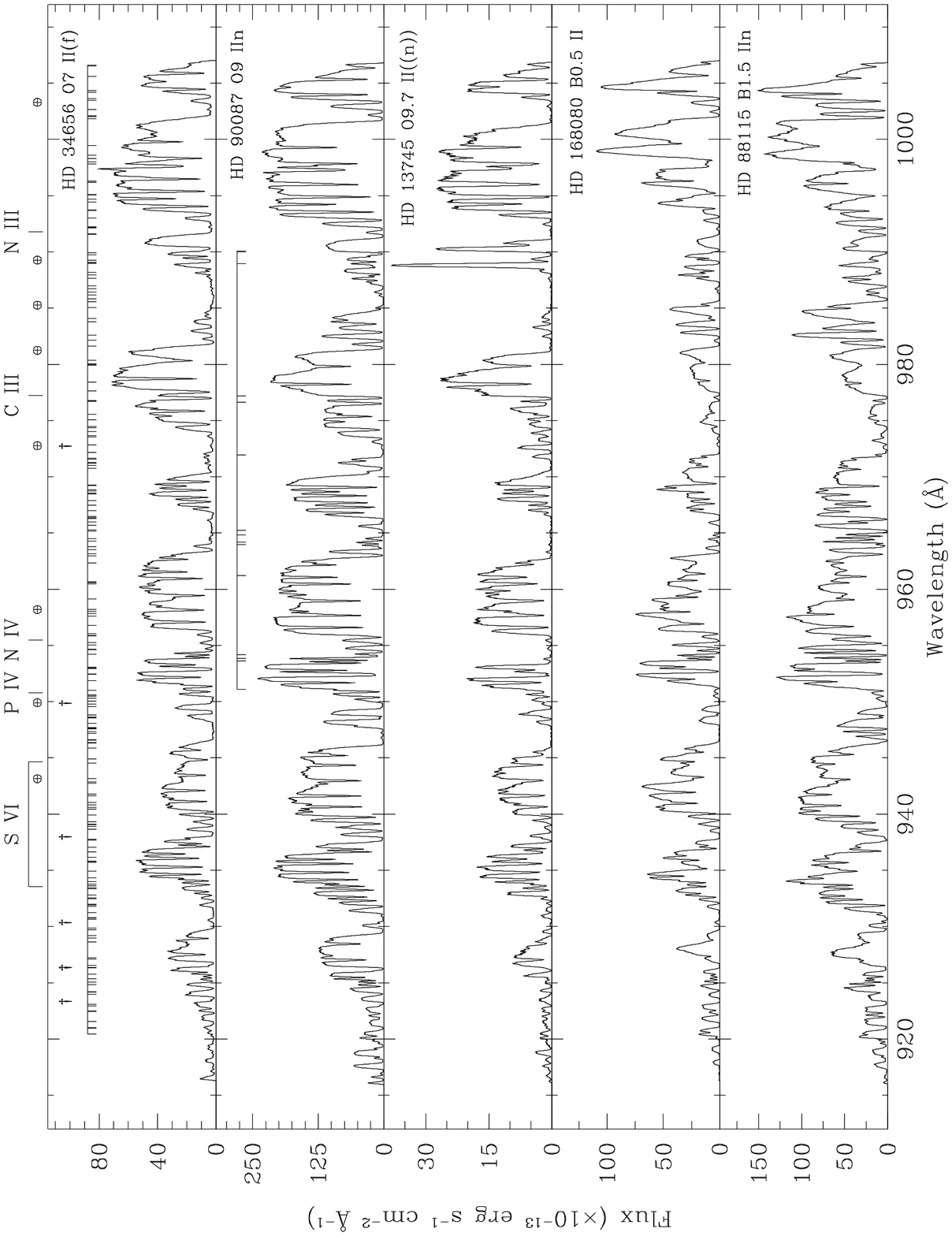}
\caption {\fuse\ spectra of bright giant stars with spectral types from O7 to B1.5
between 912 and 1012\AA. See Figure~1a for a description of the labels.}
\end{figure}

\clearpage

\begin{figure}
\figurenum{13}
\plotone{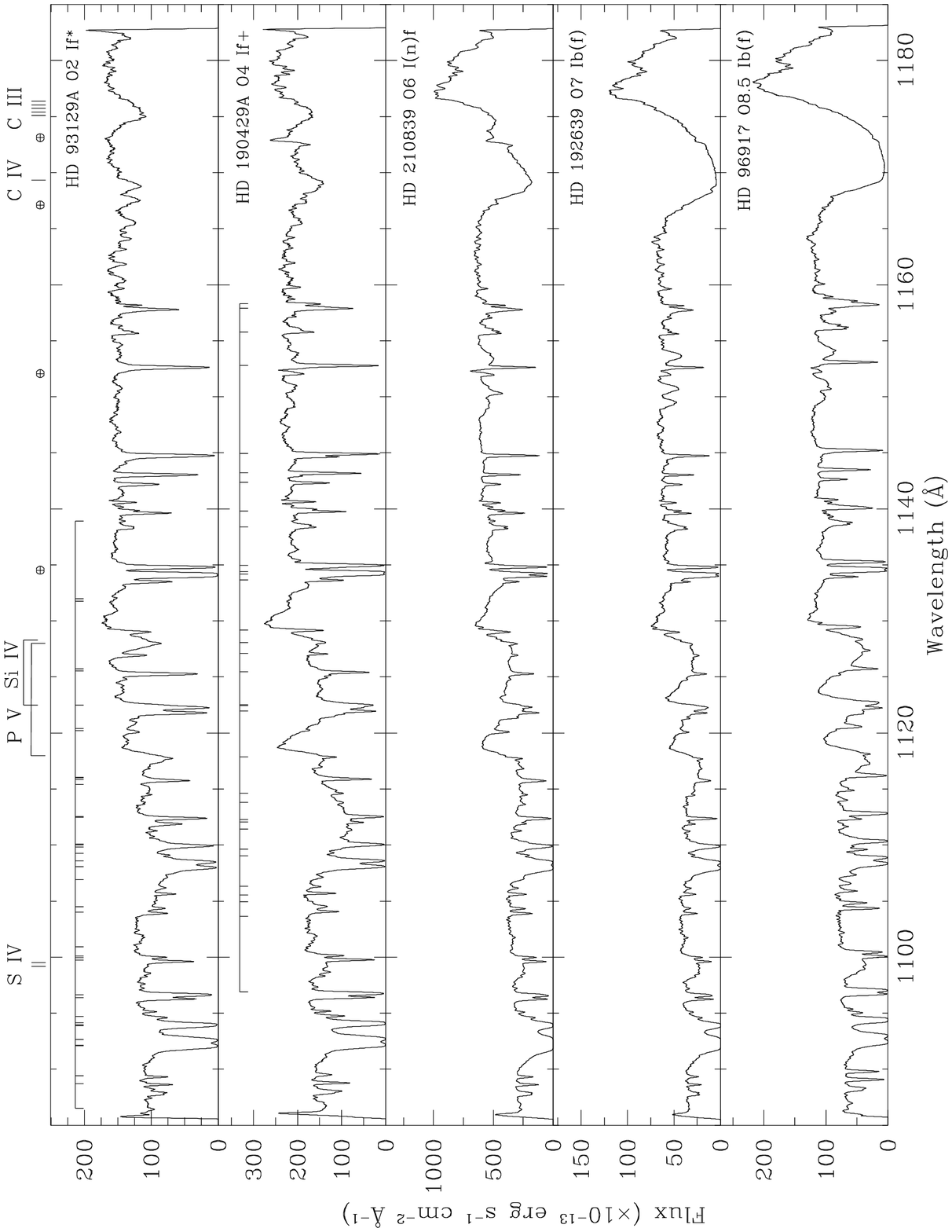}
\caption {\fuse\ spectra of supergiant stars with spectral types from (a) O2 to O8.5,
and (b) O9 to B2 between 1085 and 1185\AA. See Figure~1a for a description of the labels.}
\end{figure}

\clearpage

\begin{figure}
\figurenum{13b}
\plotone{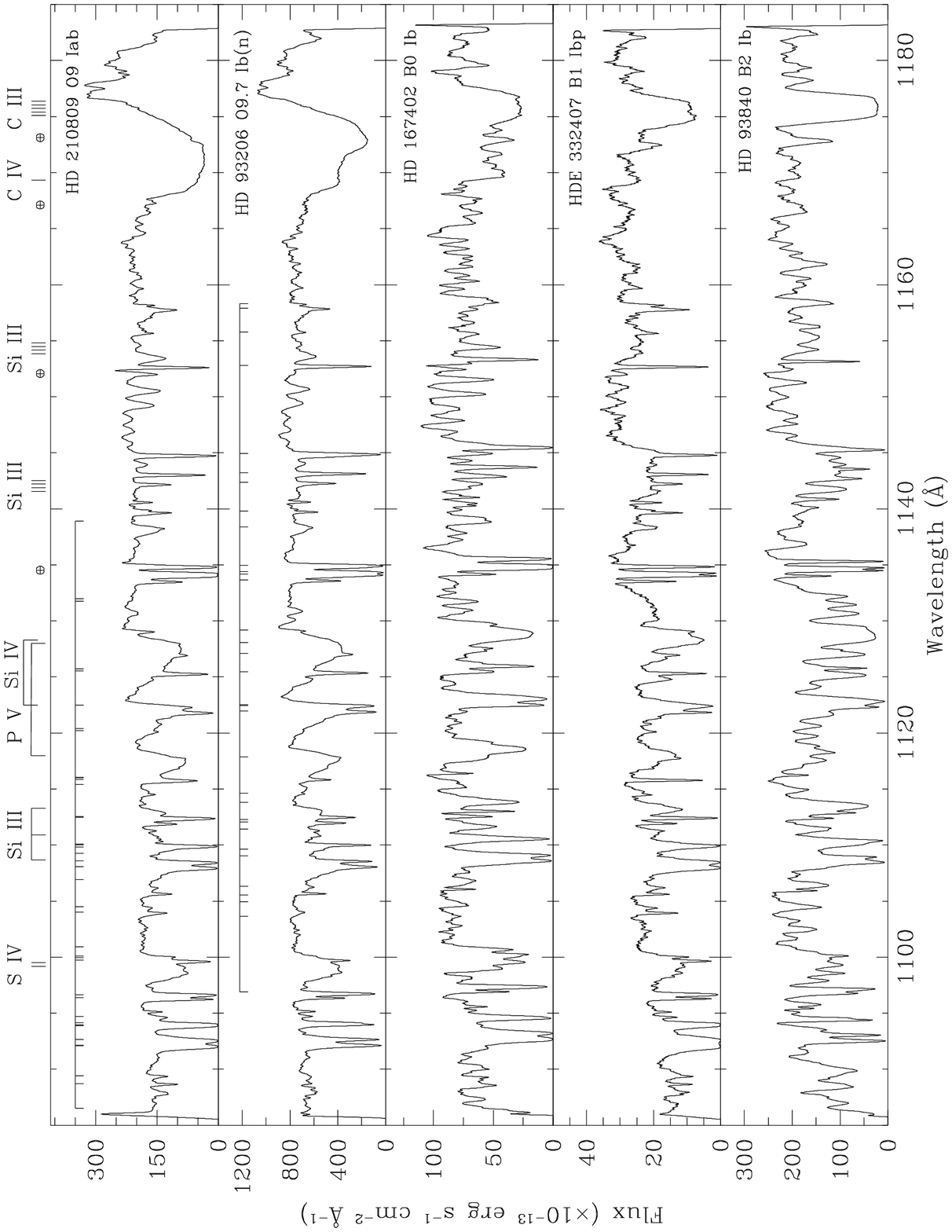}
\end{figure}

\clearpage

\begin{figure}
\figurenum{14}
\plotone{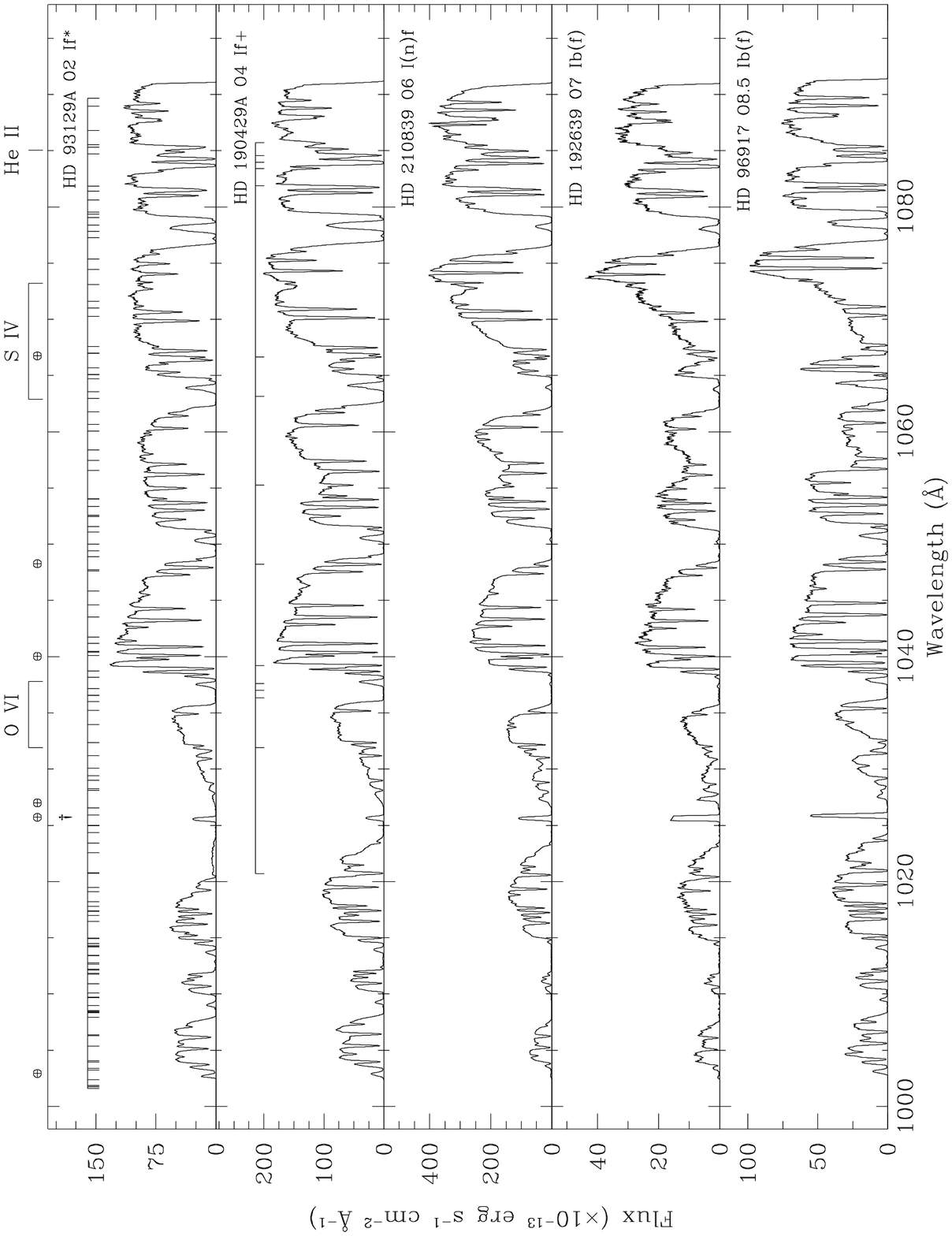}
\caption {\fuse\ spectra of supergiant stars with spectral types from (a) O2 to O8.5,
and (b) O9 to B2 between 998 and 1098\AA. See Figure~1a for a description of the labels.}
\end{figure}

\clearpage

\begin{figure}
\figurenum{14b}
\plotone{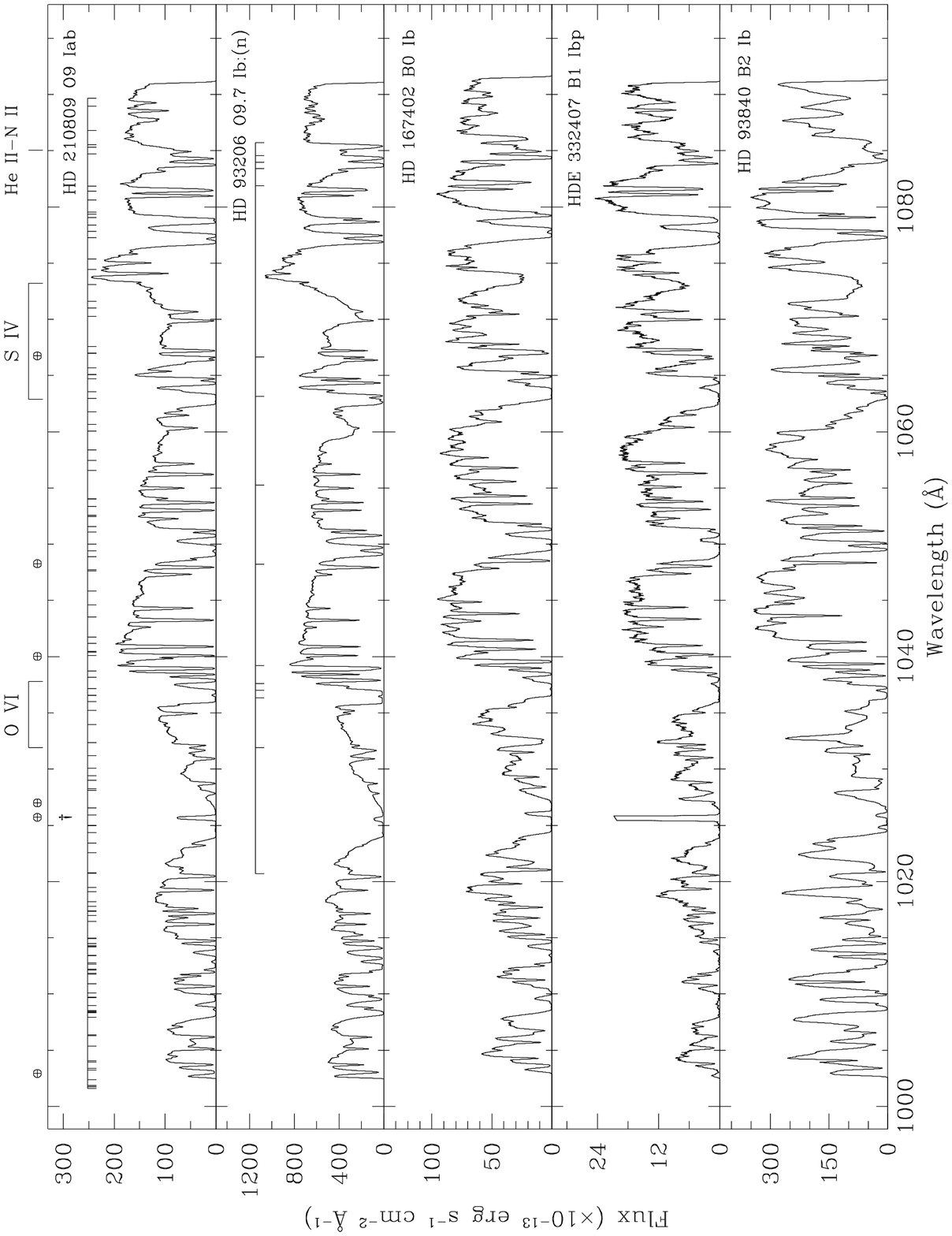}
\end{figure}

\clearpage

\begin{figure}
\figurenum{15}
\plotone{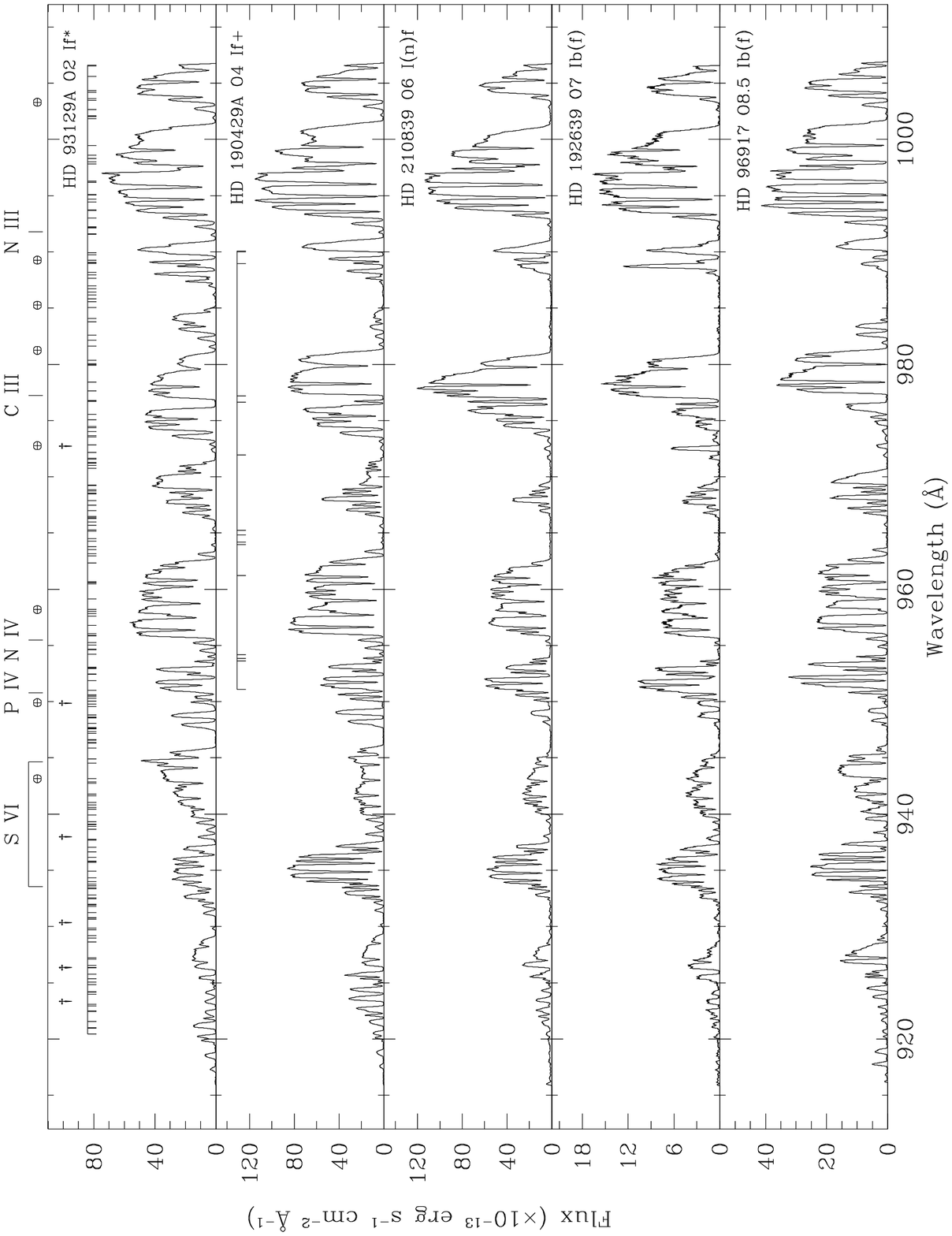}
\caption {\fuse\ spectra of supergiant stars with spectral types from (a) O2 to O8.5,
and (b) O9 to B2 between 912 and 1012\AA. See Figure~1a for a description of the labels.}
\end{figure}

\clearpage

\begin{figure}
\figurenum{15b}
\plotone{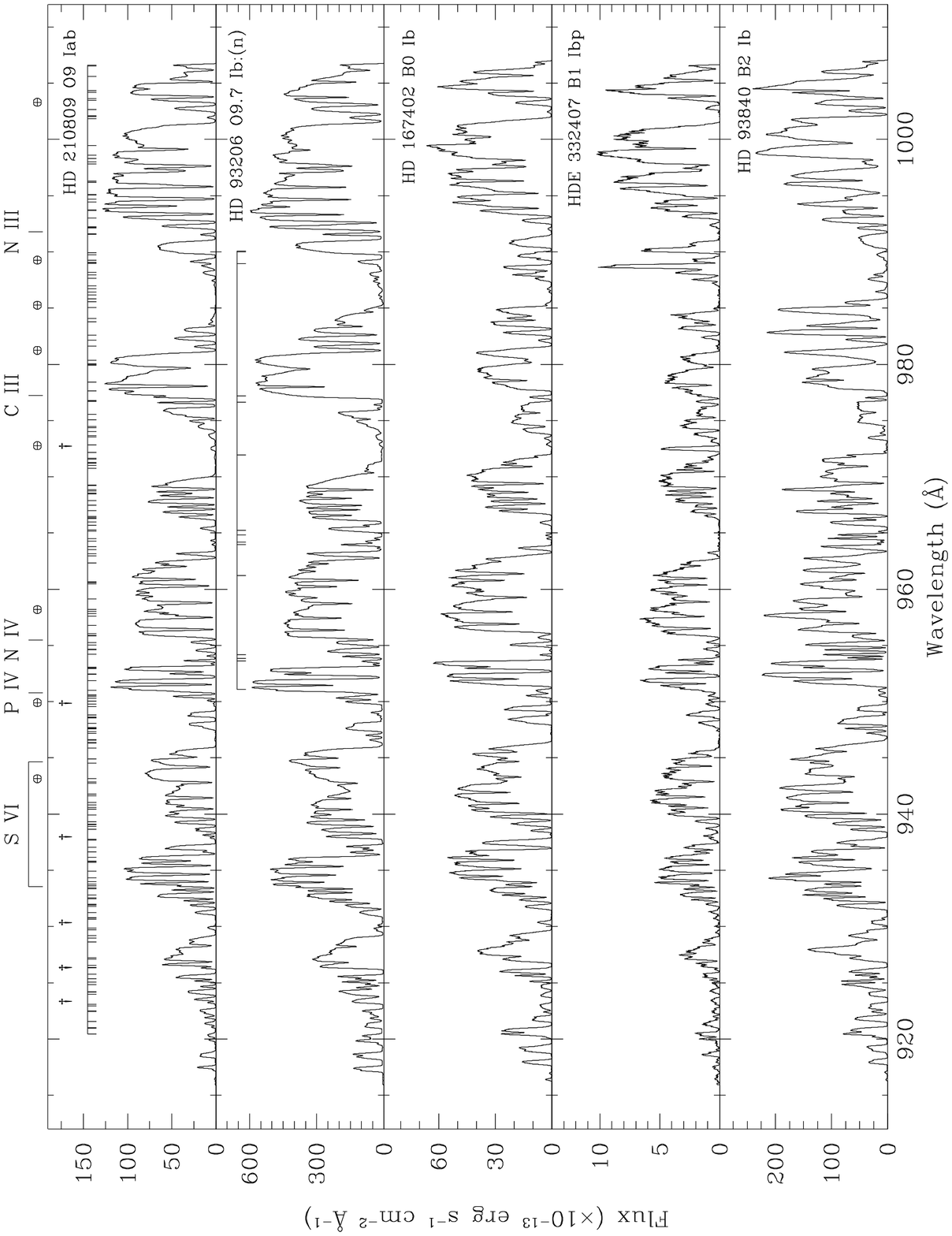}
\end{figure}

\clearpage

\begin{figure}
\figurenum{16}
\plotone{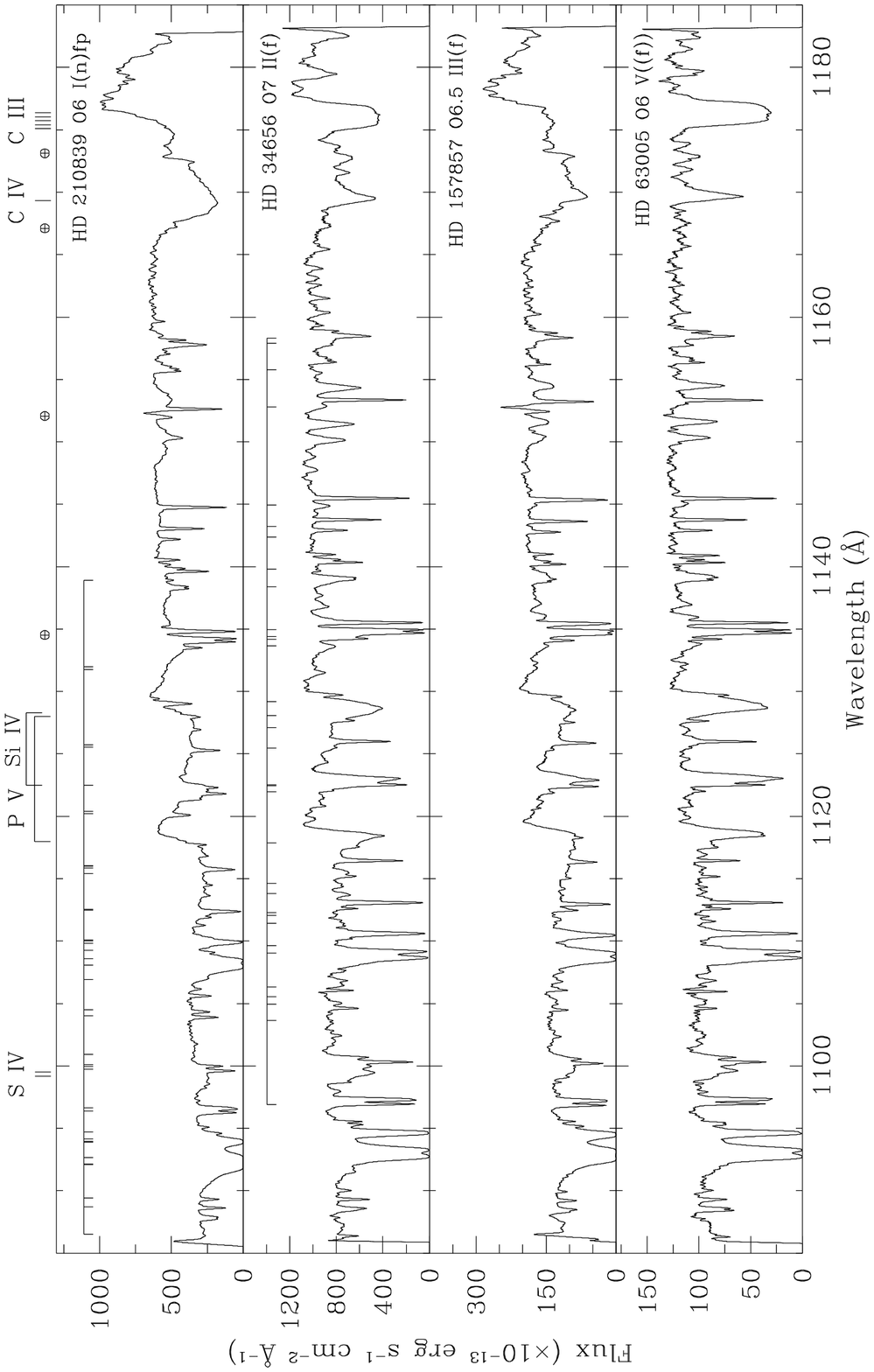}
\caption {\fuse\ spectra of four different luminosity class stars around spectral
types O6-O7 between 1085 and 1185\AA. See Figure~1a for a description of the labels.}
\end{figure}

\clearpage

\begin{figure}
\figurenum{17}
\plotone{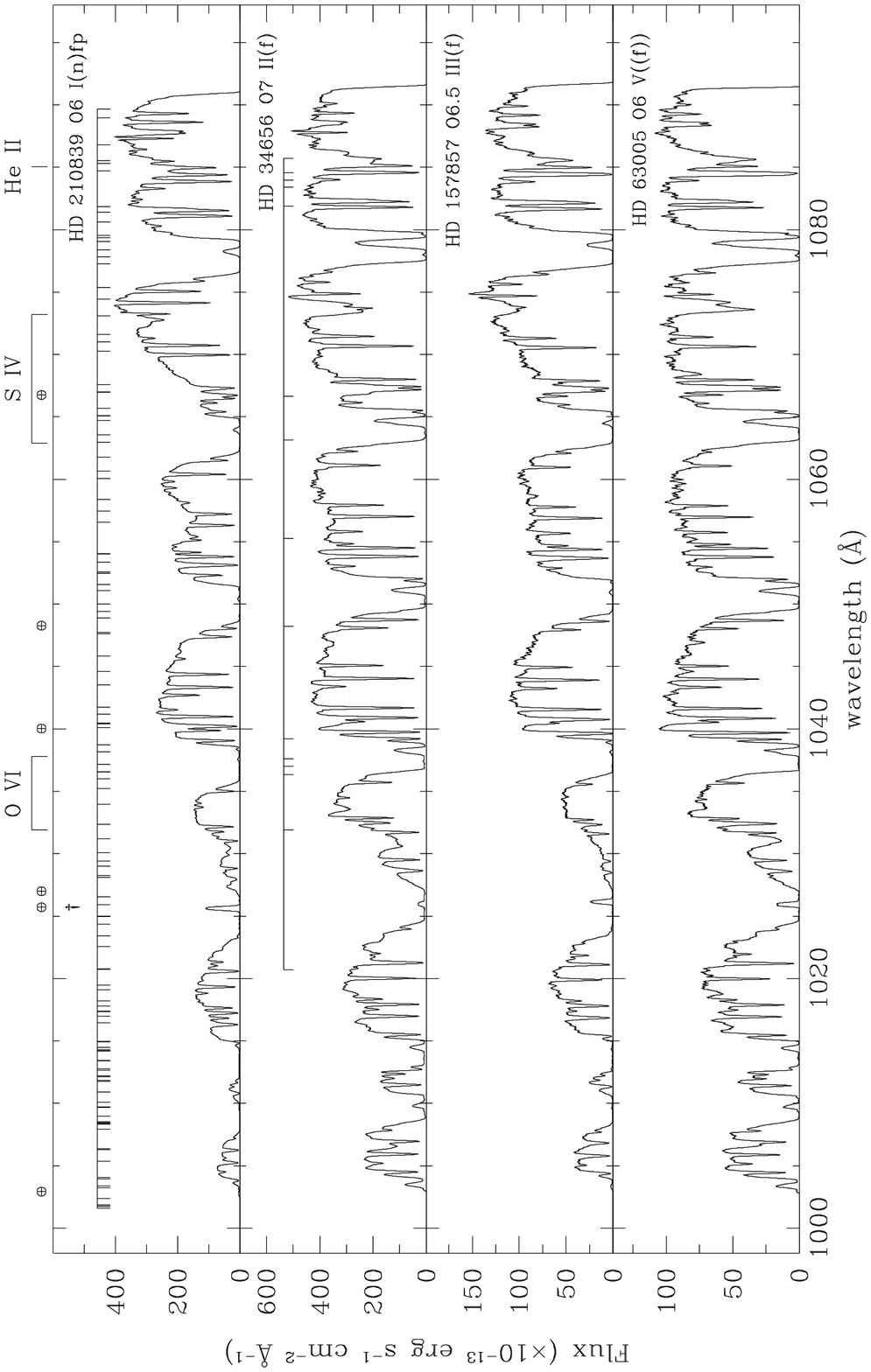}
\caption {\fuse\ spectra of four different luminosity class stars around spectral
types O6-O7 between 998 and 1098\AA. See Figure~1a for a description of the labels.}
\end{figure}

\clearpage

\begin{figure}
\figurenum{18}
\plotone{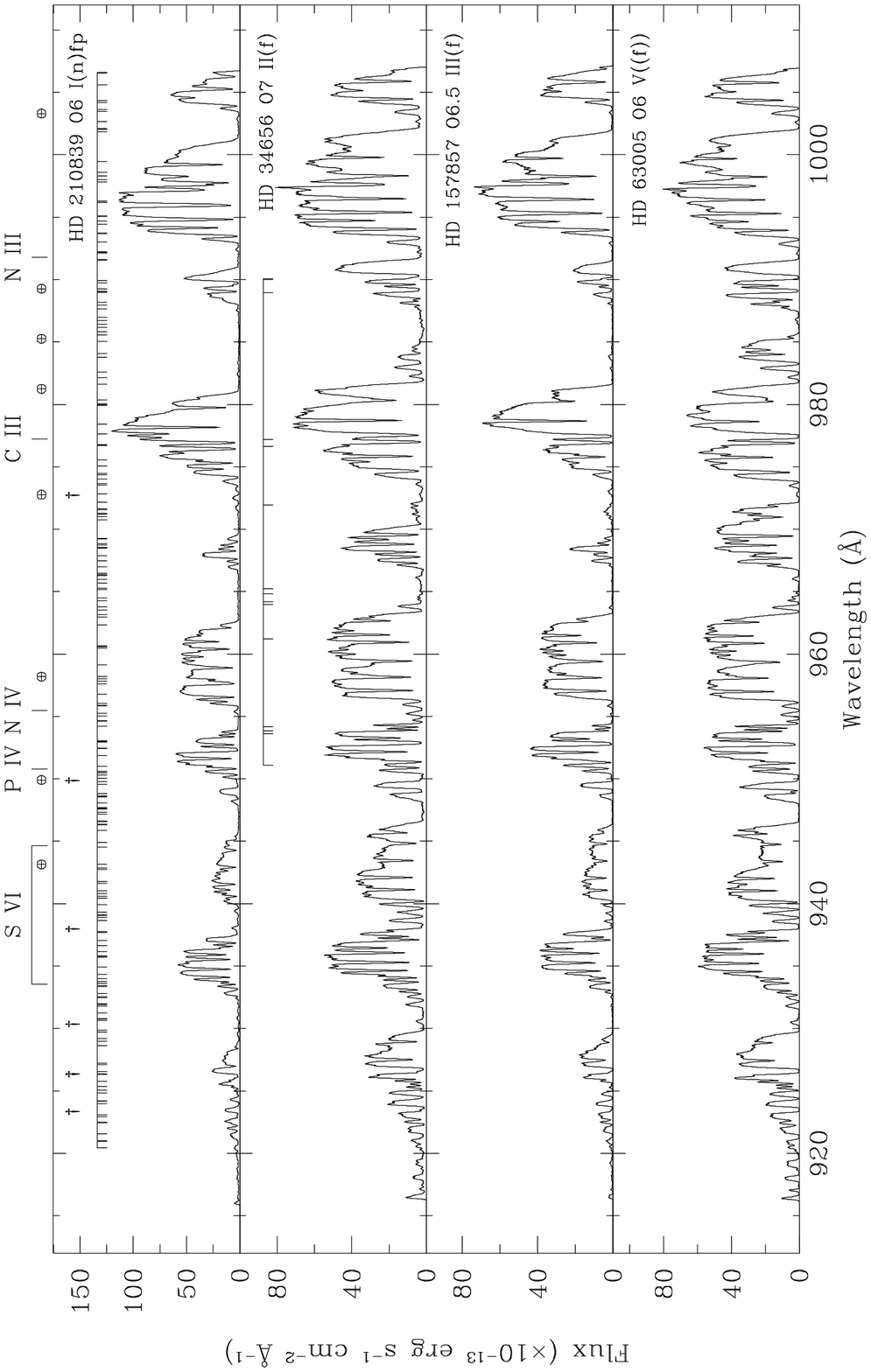}
\caption {\fuse\ spectra of four different luminosity class stars around spectral
types O6-O7 between 912 and 1012\AA. See Figure~1a for a description of the labels.}
\end{figure}

\clearpage

\begin{figure}
\figurenum{19}
\plotone{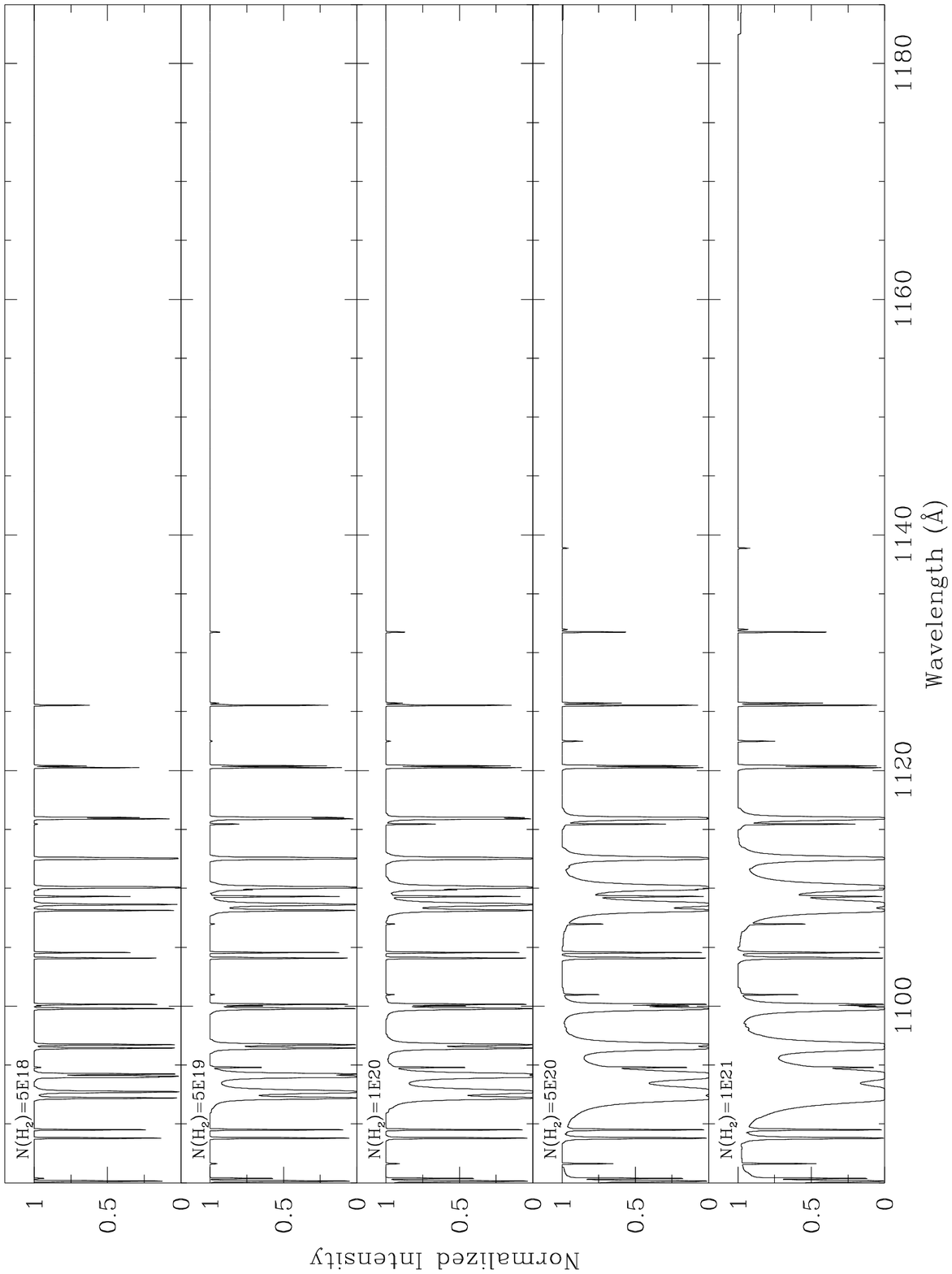}
\caption {Contribution of interstellar H$_2$ and {\ion{H}{1}} between 1085 and 1185\AA. 
The column density is indicated on left side of each panel. The adopted
N({\ion{H}{1}})/N(H$_2$) ratio is 0.5 with a H$_2$ rotational
temperature of 300K, and a line width of 5 km/s$^{-1}$. 
Note that this Figure has exactly the same scale as Figures 1, 4, 7, 10, 13,
and 16, and can therefore be used to provide approximate line identifications. See \S4 for details. }
\end{figure}

\clearpage

\begin{figure}
\figurenum{20}
\plotone{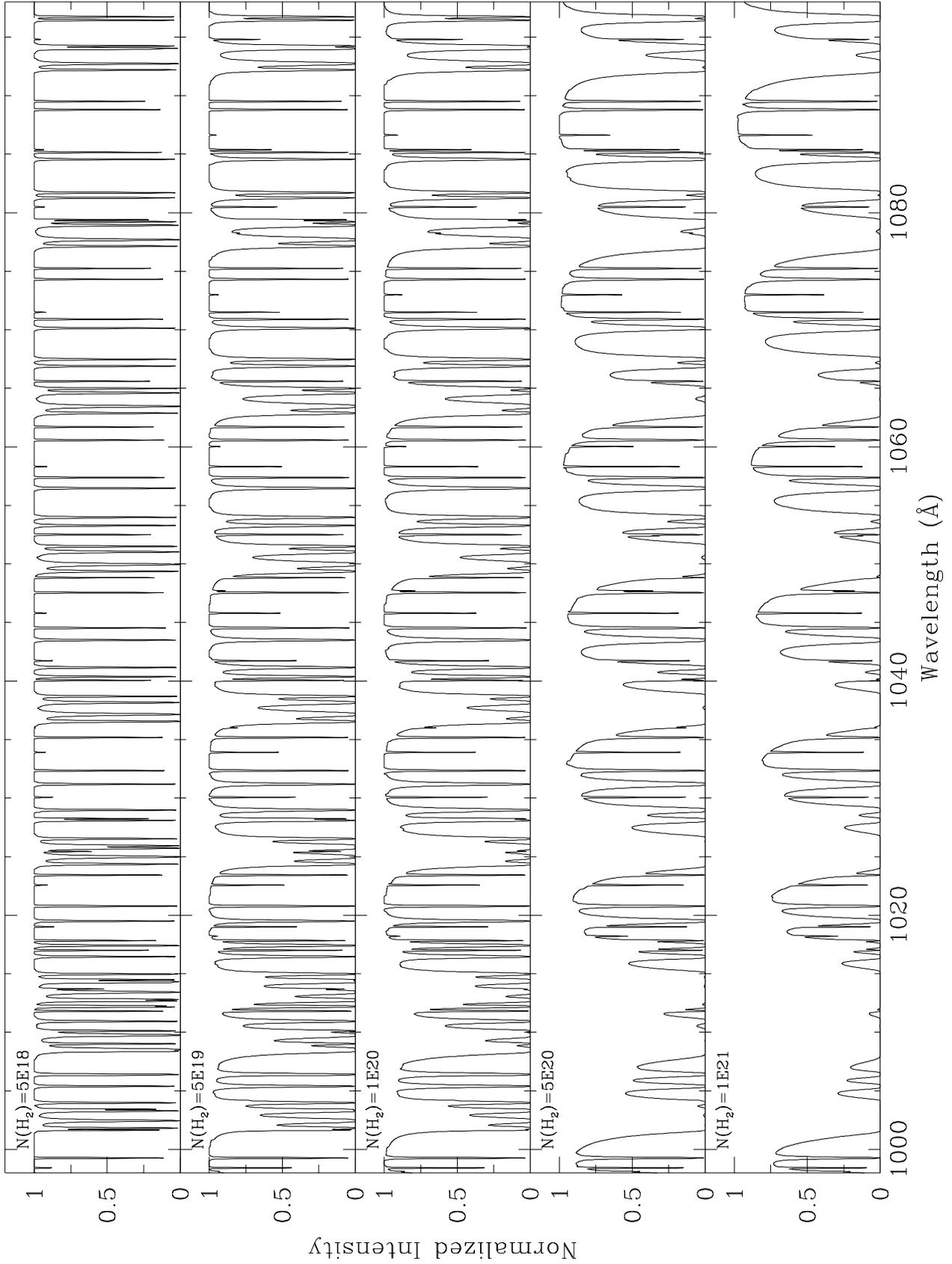}
\caption {Contribution of interstellar H$_2$ and {\ion{H}{1}} between 998 and 1098\AA. 
The model parameters are the same as Figure~19. Note that this Figure has
exactly the same scale as Figures 2, 5, 8, 11, 14, and 17, and can therefore
be used to provide approximate line identifications. See \S4 for details.}
\end{figure}

\clearpage

\begin{figure}
\figurenum{21}
\plotone{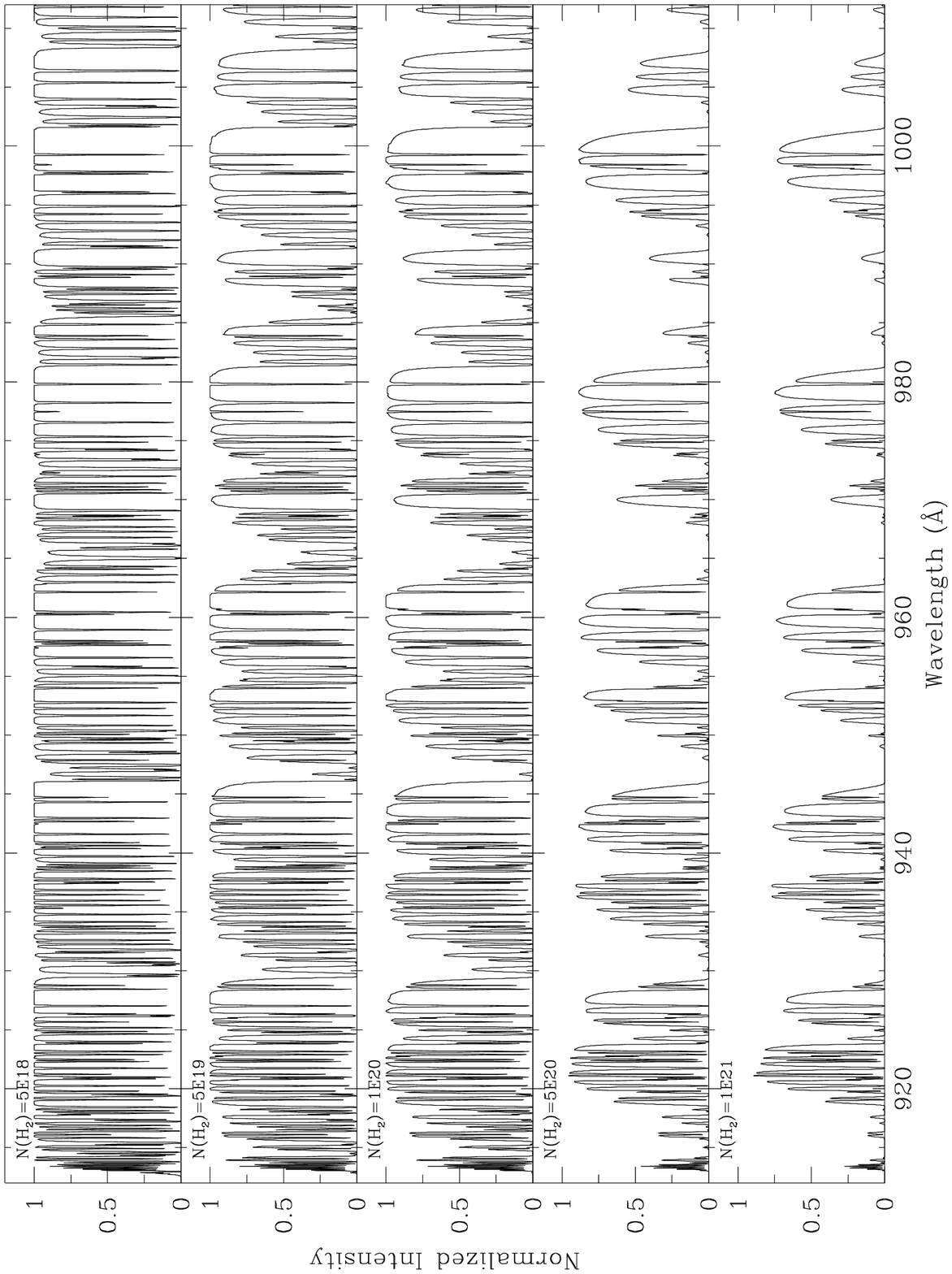}
\caption {Contribution of interstellar H$_2$ and {\ion{H}{1}} between 912 and 1012\AA.
The model parameters are the same as Figure~19.
Note that this Figure has exactly the same scale as Figures 3, 6, 9, 12, 15,
and 18, and can therefore be used to provide approximate line identifications.
See \S4 for details.}
\end{figure}

\clearpage

\begin{deluxetable}{lccrrlcrccr}
\tabletypesize{\scriptsize}
\tablecaption{Targets Included in the Atlas\label{targets}}
\tablewidth{0pt}
\tablecolumns{2}
\tablehead{ \colhead{Name} & \colhead{RA(J2000)} & \colhead{DEC(J2000)} & \colhead{$l$} & \colhead{$b$} & \colhead{Spectral Type} & \colhead{Ref.} & \colhead{{\it{V}}} & \colhead{$E(B-V)$} & \colhead{FUSE ID} & \colhead{Exp.}\\
 & \colhead{(h m s)} & \colhead{($^\circ$ $'$ $''$)} & \colhead{(deg)} & \colhead{(deg)} &  &  &  &  &  & \colhead{(sec)}}
\startdata
HD\,73 & 00 05 36.92 & +43 24 05.10 & 114.17 & $-$18.69 & B1.5 IV & 1 & 8.48 & 0.07 & P1010101 & 59 \\
HD\,13745 & 02 15 45.90 & +55 59 46.69 & 133.98 & $-$3.48 & O9.7 II((n)) & 2 & 7.83 & 0.45 & P1020404 & 4486 \\
HD\,34656 & 05 20 43.06 & +37 26 18.90 & 170.04 & +0.27 & O7 II(f) & 3 & 6.79 & 0.34 & P1011301 & 4179 \\
HD\,60369 & 07 33 01.87 & $-$28 19 32.70 & 242.68 & $-$4.30 & O9 IV & 4 & 8.15 & 0.30 & P1050201 & 7140 \\
HD\,63005 & 07 45 49.06 & $-$26 29 31.50 & 242.47 & $-$0.93 & O6 V((f)) & 4 & 9.13 & 0.27 & P1022101 & 5311 \\
\\
HD\,66788 & 08 04 08.51 & $-$27 29 09.60 & 245.43 & +2.05 & O8 V & 5 & 9.45 & 0.22 & P1011801 & 4209 \\
HD\,74711 & 08 43 47.50 & $-$46 47 56.10 & 265.73 & $-$2.61 & B1 III & 4 & 7.11 & 0.33 & P1022501 & 4684 \\
HD\,74920 & 08 45 10.52 & $-$46 02 18.89 & 265.29 & $-$1.95 & O7 IIIn & 6 & 7.53 & 0.34 & P1022601 & 4672 \\
HD\,88115 & 10 07 31.87 & $-$62 39 12.69 & 285.32 & $-$5.53 & B1.5 IIn & 4 & 8.30 & 0.16 & P1012301 & 4512 \\
HD\,89137 & 10 15 40.02 & $-$51 15 25.41 & 279.69 & +4.45 & O9.5 III(n)p & 7 & 7.98 & 0.23 & P1022801 & 4628 \\
\\
HD\,90087 & 10 22 20.85 & $-$59 45 20.00 & 285.16 & $-$2.13 & O9 IIn & 4 & 7.76 & 0.28 & P1022901 &3934 \\
HD\,91651 & 10 33 30.38 & $-$60 07 35.30 & 286.55 & $-$1.72 & O9 V:n & 2 & 8.84 & 0.29 & P1023102 & 8137 \\
HD\,93129A & 10 43 57.40 & $-$59 32 51.00 & 287.41 & $-$0.57 & O2 If* & 8 & 8.84 & 0.58 & P1170202 & 7361 \\
HD\,93146 & 10 43 59.98 & $-$60 05 11.30 & 287.67 & $-$1.05 & O6.5 V((f)) & 2 & 8.45 & 0.35 & P1023301 & 4140 \\
HD\,93204 & 10 44 32.46 & $-$59 44 30.10 & 287.57 & $-$0.71 & O5 V((f)) & 2 & 8.48 & 0.41 & P1023501 & 4664 \\
\\
HD\,93206\tablenotemark{a} & 10 44 23.05 & $-$59 59 36.10 & 287.67 & $-$0.94 & O9.7 Ib:(n) & 3 & 6.24 & 0.38 & P1023401 & 4140 \\
HD\,93250 & 10 44 45.19 & $-$59 33 54.50 & 287.51 & $-$0.54 & O3.5 V((f+)) & 8 & 7.37 & 0.48 & P1023801 & 4140 \\
HD\,93840 & 10 49 08.75 & $-$46 46 42.41 & 282.14 & +11.10 & BN1 Ib & 9 & 7.77 & 0.14 & P1012701 & 5318 \\
HD\,93843 & 10 48 37.83 & $-$60 13 25.50 & 288.24 & $-$0.90 & O5 III(f)var & 2 & 7.30 & 0.28 & P1024001 & 4140 \\
HD\,96715 & 11 07 32.93 & $-$59 57 48.80 & 290.27 & +0.33 & O4 V((f)) & 2 & 8.27 & 0.42 & P1024301 & 4597 \\
\\
HD\,96917 & 11 08 42.47 & $-$57 03 57.10 & 289.28 & +3.06 & O8.5 Ib(f) & 2 & 7.07 & 0.37 & P1024401 & 8005 \\
HD\,99890 & 11 29 05.83 & $-$56 38 38.99 & 291.75 & +4.43 & B0 IIIn & 4 & 8.28 & 0.24 & P1024601 & 4584 \\
HD\,116852 & 13 30 23.30 & $-$78 51 18.29 & 304.88 & $-$16.13 & O9 III & 10 & 8.47 & 0.22 & P1013801 & 7212 \\
HD\,121800 & 13 55 15.52 & +66 07 00.39 & 113.01 & +49.76 & B1.5 V & 11 & 9.11 & 0.08 & P1014401 & 3987 \\
HD\,152218 & 16 53 59.98 & $-$41 42 52.90 & 343.53 & +1.28 & O9.5 IV(n) & 2 & 7.61 & 0.47 & P1015402 & 9485 \\
\\
HD\,152233 & 16 54 03.45 & $-$41 47 29.10 & 343.48 & +1.22 & O6 III:(f)p & 3 & 6.59 & 0.45 & P1026702 & 4105 \\
HD\,152623 & 16 56 14.99 & $-$40 39 36.10 & 344.62 & +1.61 & O7 V(n)((f)) & 3 & 6.67 & 0.40 & P1027001 & 6056 \\
HD\,157857 & 17 26 17.40 & $-$10 59 34.00 & 12.97 & +13.81 & O6.5 III(f) & 3 & 7.78 & 0.49 & P1027501 & 3998 \\
HD\,164906 & 18 04 25.87 & $-$24 23 09.50 & 6.05 & $-$1.33 & B1 IVpe & 12 & 7.42 & 0.42 & P1027701 & 5256 \\
HD\,167402 & 18 16 18.61 & $-$30 07 29.20 & 2.26 & $-$6.39 & B0 Ib & 4 & 8.95 & 0.23 & P1016201 & 3856 \\
\\
HD\,168080 & 18 18 46.84 & $-$18 10 19.50 & 13.11 & $-$1.27 & B0.5 II & 10 & 7.61 & 0.38 & P1222701 & 5667 \\
HD\,183899 & 19 32 45.21 & $-$26 09 46.70 & 13.07 & $-$20.14 & B2 III & 13 & 9.80 & 0.16 & P1017601 & 4396 \\
HD\,190429A & 20 03 29.45 & +36 01 29.40 & 72.59 & +2.61 & O4 If+ & 3 & 6.56 & 0.51 & P1028401 & 5390 \\
HD\,192639 & 20 14 30.48 & +37 21 13.49 & 74.90 & +1.48 & O7 Ib(f) & 3 & 7.11 & 0.64 & P1162401 & 4834 \\
HD\,195965 & 20 32 25.63 & +48 12 59.10 & 85.71 & +5.00 & B0 V & 14 & 6.98 & 0.25 & P1028803 & 6440 \\
\\
HD\,210809 & 22 11 38.64 & +52 25 47.90 & 99.85 & $-$3.13 & O9 Iab & 2 & 7.54 & 0.33 & P1223102 & 7097 \\
HD\,210839 & 22 11 30.62 & +59 24 52.30 & 103.83 & +2.61 & O6 I(n)fp & 2 & 5.06 & 0.62 & P1163101 & 6050 \\
HD\,212044 & 22 20 22.70 & +51 51 39.30 & 100.64 & $-$4.35 & B1 Vpnne & 10 & 6.98 &  0.30 & P1223401 & 4544 \\
HDE\,233622 & 09 21 33.59 & +50 05 56.90 & 168.17 & +44.23 & B2 V & 15 & 10.01 & 0.03 & P1012102 & 4662 \\
HDE\,235874 & 22 32 59.78 & +51 12 56.01 & 101.97 & $-$5.93 & B3 III & 16 & 9.64 & 0.20 & P1223701 & 5509 \\
\\
HDE\,308813 & 11 37 58.52 & $-$63 18 58.80 & 294.79 & $-$1.61 & O9.5 V & 17 & 9.28 & 0.34 & P1221901 & 4257 \\
HDE\,332407 & 19 41 19.90 & +29 08 40.40 & 64.28 & +3.11 & B1 Ibp & 14 & 8.50 & 0.48 & P1222801 & 4784 \\
BD\,+35$^\circ$4258 & 20 46 12.56 & +35 32 26.41 & 77.19 & $-$4.74 & B0.5 V & 10 & 9.41 & 0.29 & P1017901 & 3965 \\
BD\,+38$^\circ$2182 & 10 49 12.91 & +38 00 14.90 & 182.16 & +62.21 & B3 V & 18 & 11.2\phn & ... & P1012801 & 12235 \\ 
BD\,+53$^\circ$2820 & 22 13 49.63 & +54 24 34.41 & 101.24 & $-$1.69 & B0 IVn & 14 & 9.95 & 0.40 & P1223201 & 5814 \\
\enddata

\tablenotetext{a}{Double spectroscopic binary.}
\tablerefs{1. \citet{wal71}; 
2. \citet{wal73}; 
3. \citet{wal72}; 
4. Garrison, Hiltner, \& Schild (1977); 
5. \citet{mac76}; 
6. \citet{vija93}; 
7. \citet{wal76}; 
8. \citet{wal02}; 
9. Walborn, Fitzpatrick, \& Nichols-Bohlin (1990);
10. Morgan, Code, \& Whitford (1955); 
11. Dworetsky, Whitelock, \& Carnochan (1982);
12. Jaschek, Conde, \& Sierra (1964); 
13. \citet{hill70}; 
14. \citet{hilt56}; 
15. \citet{ryan97};
16. \citet{cramp76}; 
17. \citet{schi70};
18. \citet{con89}.}

\end{deluxetable}

\begin{deluxetable}{lc|lc|lc}
\tabletypesize{\scriptsize}
\tablecaption{Wavelengths of Strong Stellar Lines in the FUV\label{starlines}}
\tablewidth{0pt}
\tablehead{\colhead{Line} & \colhead{$\lambda$ (\AA)} & \colhead{Line} & \colhead{$\lambda$ (\AA)} & \colhead{Line} & \colhead{$\lambda$ (\AA)}}
\startdata
Ly~$\theta$ & 923.150 & N~{\sc iii} & 991.577 & P~{\sc v} & 1117.977 \\
Ly~$\eta$ & 926.226 & Ly~$\beta$ & 1025.722\phn & Si~{\sc iv} & 1122.485 \\
Ly~$\zeta$ & 930.748 & O~{\sc vi} & 1031.926\phn & P~{\sc v} & 1128.008 \\
S~{\sc vi} & 933.378 & O~{\sc vi} & 1037.617\phn & Si~{\sc iv} & 1128.340 \\
Ly~$\varepsilon$ & 937.804 & S~{\sc iv} & 1062.662\phn & C~{\sc iii}\tablenotemark{b} & 1174.933 \\
S~{\sc vi} & 944.523 & S~{\sc iv} & 1072.974\phn &  & 1175.263 \\
Ly~$\delta$ & 949.743 &  & 1073.516\phn &  & 1175.590 \\
P~{\sc iv} & 950.657 & S~{\sc iv}\tablenotemark{a} & 1098.359\phn &  & 1175.711 \\
Ly~$\gamma$ & 972.537 &  & 1098.930\phn &  & 1175.987 \\
C~{\sc iii} & 977.020 &  & 1099.481\phn &  & 1176.370 \\
N~{\sc iii} & 991.511 &  & 1100.053\phn & \\
\enddata

\tablenotetext{a}{The S~{\sc iv} multiplet is centered at 1099.4\,\AA.}
\tablenotetext{b}{The C~{\sc iii} multiplet is centered at 1175.6\,\AA.}

\end{deluxetable}

\begin{deluxetable}{ll|ll|ll|ll}
\tabletypesize{\scriptsize}
\tablecaption{Important Interstellar Lines in the FUV\label{ismlines}}
\tablewidth{0pt}
\tablehead{\colhead{Ion}                   & 
           \colhead{$\lambda_{lab}$ (\AA)} & 
	   \colhead{Ion}                   & 
	   \colhead{$\lambda_{lab}$ (\AA)} & 
	   \colhead{Ion}                   & 
	   \colhead{$\lambda_{lab}$ (\AA)} & 
	   \colhead{Ion}                   & 
	   \colhead{$\lambda_{lab}$ (\AA)} }
\startdata
O~{\sc i} & \phn950.885 & O VI & 1031.926 & C~{\sc i} & 1104.941 & C~{\sc i} & 1128.081 \\
N~{\sc i} & \phn953.415 & C~{\sc ii} & 1036.337 & C~{\sc i} & 1105.529 & C~{\sc i} & 1129.195 \\
N~{\sc i} & \phn953.655 & C~{\sc ii}$^{\ast}$ & 1037.018 & C~{\sc i} & 1106.316 & Fe~{\sc ii} & 1133.665 \\
N~{\sc i} & \phn953.970 & O VI & 1037.617 & C~{\sc i} & 1109.041 & N~{\sc i} & 1134.165 \\
P~{\sc ii} & \phn961.041 & O~{\sc i} & 1039.230 & C~{\sc i} & 1109.633 & N~{\sc i} & 1134.415 \\
P~{\sc ii} & \phn963.801 & Ar~{\sc i} & 1048.220 & C~{\sc i} & 1111.420 & N~{\sc i} & 1134.980 \\
N~{\sc i} & \phn963.990 & Fe~{\sc ii} & 1055.270 & Fe~{\sc ii} & 1112.048 & C~{\sc i} & 1138.384 \\
N~{\sc i} & \phn964.626 & Ar~{\sc i} & 1066.660 & C~{\sc i} & 1112.268 & C~{\sc i} & 1139.792\tablenotemark{a} \\
N~{\sc i} & \phn965.041 & Fe~{\sc ii} & 1063.176 & C~{\sc i} & 1113.793 & Fe~{\sc ii} & 1142.366 \\
O~{\sc i} & \phn971.738\tablenotemark{a} & Fe~{\sc ii} & 1081.875 & C~{\sc i} & 1114.626 & Fe~{\sc ii} & 1143.226 \\
O~{\sc i} & \phn976.448 & Fe~{\sc ii} & 1083.420 & C~{\sc i} & 1117.865 & Fe~{\sc ii} & 1144.938 \\
C~{\sc iii} & \phn977.020 & N~{\sc ii} & 1083.994 & Fe~{\sc ii} & 1121.975 & P~{\sc ii} & 1152.818 \\
O~{\sc i} & \phn988.773\tablenotemark{a} & N~{\sc ii}$^{\ast}$ & 1084.584\tablenotemark{a} & C~{\sc i} & 1122.437\tablenotemark{a} & C~{\sc i} & 1155.810 \\
N~{\sc iii} & \phn989.799 & N~{\sc ii}$^{\ast\ast}$ & 1085.710\tablenotemark{a} & Fe~{\sc iii} & 1122.524\tablenotemark{a} & C~{\sc i} & 1157.910 \\
Si~{\sc ii} & \phn989.873 & Fe~{\sc ii} & 1096.890 & Fe~{\sc ii} & 1125.448 & C~{\sc i} & 1158.324 \\
Si~{\sc ii} & 1020.699 & C~{\sc i} & 1103.629 & Fe~{\sc ii} & 1127.098 & \\
\enddata

\tablenotetext{a}{Blend of multiple lines.}
\tablecomments{Asterisks indicate a transition from excited levels.}

\end{deluxetable}

\begin{deluxetable}{lcc|lcc}
\tabletypesize{\scriptsize}
\tablecaption{Strong Airglow Lines\tablenotemark{a}~ in {\fuse}
	      Spectra\label{aglines}}
\tablewidth{0pt}
\tablewidth{0pt}
\tablehead{\colhead{Line}             &
	   \colhead{$\lambda$ (\AA)}  &
	   \colhead{Note}             &
	   \colhead{Line}             &
	   \colhead{$\lambda$ (\AA)}  &
	   \colhead{Note}}
\startdata
N$_2$    &  943 & b              & O~{\sc i}  & 1027     & Multiplet  \\
Ly~$\delta$ &  949.743 &            & O~{\sc i}  & 1040     & Multiplet  \\
N$_2$    &  958 &               &Ar~{\sc i} & 1048.220 &            \\
Ly~$\gamma$ &  972.537 &            & Ar~{\sc i} & 1066.660 &            \\
N$_2$    &  981 &               &N~{\sc i}  & 1134.5   & b; Multiplet \\
N$_2$    &  985 &  b             &O~{\sc i}  & 1152 & b; $^1D - ^1D^0$ \\
O~{\sc i}   &  989     & b; Multiplet &N~{\sc i}  & 1167 & $^2D^0 - ^2F$ \\
N$_2$    & 1003 &               &O~{\sc i}  & 1173 & b; $^1D - ^3D^0$ \\
Ly~$\beta$  & 1025.722 & b          &           &      &               \\
\enddata

\tablenotetext{a}{From \citealt{feld01}.}
\tablecomments{b: Bright transition that can cause a defect.}
\end{deluxetable}


\clearpage

\begin{thebibliography}{}

\bibitem[Barnstedt et al.(1999)]{barn99}
Barnstedt, J. et al. 1999, \aaps, 134, 561

\bibitem[Barnstedt et al.(2000)]{barn00} 
Barnstedt, J., Gringel, W., Kappelmann, N., \& 
Grewing, M. 2000, \aaps, 143,193

\bibitem[Conlon et al.(1989)]{con89}
Conlon, E. S., Brown, P. J. F., Dufton, P. L., \&
Keenan, F. P. 1989, \aap, 224, 65

\bibitem[Crampton et al.(1976)]{cramp76} 
Crampton, D., Bernard, D., Harris, B. L., \& Thackeray, A. D. 1976, 
\mnras, 176, 683

\bibitem[Davidsen et al.(1992)]{dav92} 
Davidsen, A. F., et al. 1992, \apj, 392, 264

\bibitem[Diplas \& Savage(1994)]{dip94}
Diplas, A., \& Savage, B. D. 1994, \apj, 427, 274

\bibitem[Dworetsky et al.(1982)]{dwor82}
Dworetsky, M. M., Whitelock, P. A., \& Carnochan, D. J. 1982, \mnras, 201, 901

\bibitem[Feldman et al.(2001)]{feld01}
Feldman, P. D., Sahnow, D. J., Kruk, J. W., Murphy, E. M., \& Moos, H. W. 2001,
\jgr, 106, 8119

\bibitem[Garrison et al.(1977)]{garr77} 
Garrison, R. F., Hiltner, W. A., \& Schild, R. E. 1977, \apjs, 35, 111

\bibitem[Garc\'{\i}a \& Mermilliod(2001)]{gar01}
Garc\'{\i}a, B., \& Mermilliod, J. C. 2001, \aaps, 368, 122

\bibitem[Hill(1970)]{hill70} 
Hill, P. W. 1970, \mnras, 150, 23

\bibitem[Hiltner(1956)]{hilt56} 
Hiltner, W. A. 1956, \apjs, 2, 389

\bibitem[Hurwitz et al.(1998)]{hur98}
Hurwitz, M., Bowyer, S., Bristol, R., Dixon, W. V. D., Dupuis, J., 
Edelstein, J., Jelinsky, P., Sasseen, T. P., \& Siegmund, O. 1998,
\apj, 500, L1

\bibitem[Jaschek et al.(1964)]{jas64} 
Jaschek, C., Conde, H., \& de Sierra, A. C. 1964, PLPla, 28, 1

\bibitem[Jenkins et al.(1996)]{jen96}
Jenkins, E. B., Reale, M. A., Zucchino, P. M., \& Sofia, U. J. 1996,
\apss, 239, 315

\bibitem[Lehner et al.(2001)]{leh01}
Lehner, N., Fullerton, A. W., Sembach, K. R., Massa, D. L., \& Jenkins, E. B.
2001, \apj, 556, L103

\bibitem[Leitherer et al.(1999)]{star99} 
Leitherer, C., Schaerer, D., Goldader, J. D., Gonz\'alez-Delgado, R. M., 
Robert, C., Kune, D. F., de Mello, D. F., Devost, D., \& Heckman, T. M. 
1999, \apjs, 123, 3

\bibitem[MacConnell \& Bidelman(1976)]{mac76} 
MacConnell, D. J., \& Bidelman, W. P. 1976, \aj, 81, 225

\bibitem[MacFarlane et al.(1993)]{mac93}
MacFarlane, J. J., Waldron, W. L., Corcoran, M. F., Wolff, M. J.,
Wang, P., \& Cassinelli, J. P. 1993, \apj, 419

\bibitem[Morgan et al.(1955)]{morg55} 
Morgan, W. W., Code, A. D., \& Whitford, A. E., 1955, \apjs, 2, 41

\bibitem[Morton \& Dinerstein(1976)]{mort76} 
Morton, D. C., \& Dinerstein, H. L. 1976, \apj, 204, 1

\bibitem[Morton \& Underhill(1977)]{mort77} 
Morton, D. C., \& Underhill, A. B. 1977, \apjs, 33, 83

\bibitem[Morton(1991)]{mort91} 
Morton, D. C. 1991, \apjs, 77, 119

\bibitem[Moos et al.(2000)]{moos00} 
Moos, H. W., et al. 2000, \apj, 538, L1

\bibitem[Rogerson \& Ewell(1985)]{rog85}
Rogerson, J. B., \& Ewell, N. W. 1985, \apjs, 58, 265

\bibitem[Rogerson et al.(1973)]{rog73}
Rogerson, J. B., Spitzer, L., Drake, J. F., Dressler, K., Jenkins, E. B., 
Morton, D. C., \& York, D. G. 1973, \apj, 181, L97

\bibitem[Ryans et al.(1997)]{ryan97}
Ryans, R. S. I., Keenan, F. P., Sembach, K. R., \& Davies, R. D. 1997, \mnras, 289, 986

\bibitem[Sahnow et al.(2000)]{sah00} 
Sahnow, D. J., et al. 2000, \apj, 538, L7

\bibitem[Schild(1970)]{schi70} 
Schild, R. E. 1970, \apj, 161, 855

\bibitem[Sembach(1999)]{sem99} 
Sembach, K. R. 1999, in ASP Conf. Ser. 166, High Velocity Clouds, 
ed. B. K. Gibson \& M. E. Putman (San Francisco; ASP), 243

\bibitem[Snow \& Morton(1976)]{snow76} 
Snow, T. P., \& Morton, D. C. 1976, \apjs, 32, 429

\bibitem[Snow \& Jenkins(1977)]{snow77}
Snow, T. P., \& Jenkins, E. B. 1977, \apjs, 33, 269

\bibitem[Stalio \& Selvelli(1975)]{sta75} 
Stalio, R., \& Selvelli, P. L. 1975, \aaps, 21, 241 

\bibitem[Taresch et al.(1997)]{tar97}
Taresch, G., Kudritzki, R. P., Hurwitz, M., Bowyer, S., Pauldrach, A. W. A., 
Puls, J., Butler, K., Lennon, D. J., \& Haser, S. M. 1997, \aap, 321, 531 

\bibitem[Vijapurkar \& Drilling(1993)]{vija93} 
Vijapurkar, J., \& Drilling, J. S. 1993, \apjs, 89, 293

\bibitem[Walborn(1971)]{wal71} 
Walborn, N. R. 1971, \apjs, 23, 257

\bibitem[Walborn(1972)]{wal72} 
Walborn, N. R. 1972, \aj, 77, 312

\bibitem[Walborn(1973)]{wal73} 
Walborn, N. R. 1973, \aj, 78, 1067

\bibitem[Walborn(1976)]{wal76} 
Walborn, N. R. 1976, \apj, 205, 419

\bibitem[Walborn et al.(1985)]{walNASA} 
Walborn, N. R., Nichols-Bohlin, J., \& Panek, R. J. 1985, 
International Ultraviolet Explorer Atlas of O-Type Spectra from
1200 to 1900~{\AA} (NASA RP-1155)

\bibitem[Walborn et al. (1990)]{wal90} 
Walborn, N. R., Fitzpatrick, E. L., \& Nichols-Bohlin, J. 1990, \pasp, 102, 543

\bibitem[Walborn \& Bohlin(1996)]{wal96} 
Walborn, N. R., \& Bohlin, R. C. 1996, \pasp, 108, 477

\bibitem[Walborn et al. (2002a)]{wal02} 
Walborn, N. R., Howarth. I. D., Lennon, D. J., Massey, P., Moffat, A.F.J., 
Skalkowski, G., Morrell, N. I., Drissen, L., Parker, J. W. 2002a, \aj, 123, 2754

\bibitem[Walborn et al. (2002b)]{mcatlas}
Walborn, N. R., Fullerton, A. W., Crowther, P. A., Bianchi, L., 
Hutchings, J. B., Pellerin, A., Sonneborn, G., \& Willis, A.
2002b, \apjs, in press.

\bibitem[Werner \& Rauch(2001)]{wer01}
Werner, K., \& Rauch, T. 2001, in ASP Conf. Ser. 242,
Eta Carinae and Other Mysterious Stars: the Hidden Opportunities of
Emission Spectroscopy, ed. T. R. Gull, S. Johansson, \& K. Davidson
(San Francisco; ASP), 229

\end{thebibliography}
\end{document}